\documentclass[11pt]{article}
\pdfoutput=1
\newcommand{\Comment}[1]{{}}
\usepackage{amsfonts,amsthm,amsmath,amssymb,slashed}
\usepackage[textwidth = 430 pt, textheight = 630 pt]{geometry}
\usepackage{color}

\Comment{\usepackage{color}
\definecolor{MyDarkBlue}{rgb}{0.15,0.15,0.45}
\usepackage[linktocpage=true]{hyperref}
\hypersetup{
colorlinks=true,
citecolor=MyDarkBlue,
linkcolor=MyDarkBlue,
urlcolor=MyDarkBlue,
pdfauthor={Horatiu Nastase and Jacob Sonnenschein},
pdftitle={$T\bar T$-like deformations of free pion action applied to the Heisenberg model and chiral perturbation theory},
pdfsubject={hep-th}
}
\usepackage[utf8]{inputenc}
\usepackage[numbers,sort&compress]{natbib}
\usepackage{hypernat}}
\usepackage{graphicx}
\usepackage{cite}
\usepackage[colorlinks=true]{hyperref}
\hypersetup{
  allcolors = blue,
}

\newcommand\ignore[1]{}
\def\one{{\,\hbox{1\kern-.8mm l}}}

\def\Tr{{\rm Tr\, }}

\def\a{\alpha}\def\b{\beta}

\def\d{\partial}

\def\Tr{\mathop{\rm Tr}\nolimits}

\newcommand{\Cset}{{\,\,{{{^{_{\pmb{\mid}}}}\kern-.45em{\mathrm C}}}}}

\newcommand{\be}{\begin{equation}}
\newcommand{\bea}{\begin{eqnarray}}

\newcommand{\ee}{\end{equation}}
\newcommand{\eea}{\end{eqnarray}}

\newcommand{\nn}{\nonumber}

\newcommand{\half}{\frac{1}{2}}

\newcommand{\CR}{\nn\cr}
\newcommand{\pa}{\partial}

\begin{document}

\renewcommand{\thefootnote}{\fnsymbol{footnote}}

\makeatletter
\@addtoreset{equation}{section}
\makeatother
\renewcommand{\theequation}{\thesection.\arabic{equation}}

\rightline{}
\rightline{}




\begin{center}
{\LARGE \bf{\sc A $T\bar T$-like deformation of the Skyrme model 
and  the Heisenberg model of nucleon-nucleon scattering}}
\end{center}
 \vspace{1truecm}
\thispagestyle{empty} \centerline{
{\large \bf {\sc Horatiu Nastase${}^{a}$}}\footnote{E-mail address: \Comment{\href{mailto:horatiu.nastase@unesp.br}}{\tt horatiu.nastase@unesp.br}}
{\bf{\sc and}}
{\large \bf {\sc Jacob Sonnenschein${}^{b,}$}}\footnote{E-mail address: \Comment{\href{mailto:cobi@tauex.tau.ac.il}}{\tt cobi@post.tau.ac.il}}
                                                        }

\vspace{.5cm}

\centerline{{\it ${}^a$Instituto de F\'{i}sica Te\'{o}rica, UNESP-Universidade Estadual Paulista}}
\centerline{{\it R. Dr. Bento T. Ferraz 271, Bl. II, Sao Paulo 01140-070, SP, Brazil}}

\vspace{.3cm}

\centerline{{\it ${}^b$School of Physics and Astronomy,}}
 \centerline{{\it The Raymond and Beverly Sackler Faculty of Exact Sciences, }} 
 \centerline{{\it Tel Aviv University, Ramat Aviv 69978, Israel}}

\vspace{1truecm}

\thispagestyle{empty}

\centerline{\sc Abstract}

\vspace{.4truecm}

\begin{center}
\begin{minipage}[c]{380pt}
{\noindent 
The Skyrme model, though it  admits correctly a wide range of static properties of the nucleon, does not seem to reproduce 
properly the scattering behavior of nucleons at high energies. In this paper we present a $T\bar T$-like deformation 
of it, inspired  by a  1+1 dimensional model, in which  boosted nucleons behave like shock waves. 
The scattering of the  latter     saturates the Froissart bound. 
We start by showing  that 1+1 dimensional $T\bar T$ deformations of the free 
abelian pion action are in fact  generalizations of the old  Heisenberg model
for nucleon-nucleon scattering, yielding the same saturation of the Froissart bound. 
We then  deform the strong coupling limit of the  bosonized  action of  multi-flavor QCD in two dimensions  
using the $T\bar T$ deformation of the WZW action  with a mass term. We derive the  classical soliton solution 
that corresponds to the nucleon, determine its mass and  discuss  its transformation  into a shock-wave upon boosting.
We uplift this action into a 3+1 dimensional $T\bar T$-like deformation of the Skyrme action.
We compare this deformed action to that of chiral perturbation theory.
 A possible holographic gravity dual interpretation is explored.
}
\end{minipage}
\end{center}

\vspace{.5cm}

\setcounter{page}{0}
\setcounter{tocdepth}{2}

\newpage

\renewcommand{\thefootnote}{\arabic{footnote}}
\setcounter{footnote}{0}

\linespread{1.1}
\parskip 4pt


\tableofcontents

\section{Introduction}

An important question that has  been intensively investigated for about a half a century  is what happens when nucleons     (or, more generally, hadrons)
 collide at very high energy? 
 It is well known  that the asymptotic  total cross section is 
bounded by the Froissart bound \cite{Froissart:1961ux,Lukaszuk:1967zz}
\be
\sigma_{\rm tot}(\tilde s)\leq C\ln^2\frac{\tilde s}{s_0}\;,\;\;\; C\leq \frac{\pi}{m^2}\;,
\ee
where $\tilde s$ is the Mandelstam kinematical variable, $s_0$ a constant  and  $m$ is the  mass of the lightest particle that can be exchanged by the
scattering projectiles.
But before the  Froissart bound, and in fact  even  QCD were  derived, Heisenberg proposed a simple semi-classical model \cite{Heisenberg1952}  that  deals 
with the regime of  high-energy nucleon scattering, leading to the saturation of the  bound. The model, which   is defined in terms of a single 
 pion field, described by  a massive DBI scalar action in 3+1 dimensions takes the following form:
\be
S=\int d^{3+1}x\,l^{-4}\left[1-\sqrt{1+l^4[(\d_\mu\phi)^2+m^2\phi^2]}\right].\label{Heisaction}
\ee
 This action is    then 
reduced to 1+1 dimensions (time and the direction of propagation).  

The colliding Lorentz-contracted nuclei and the pion field around them were described by classical shockwave solutions of this 
action.  Invoking  some simple assumptions it was shown that the total cross section $\sigma_{\rm tot}(s)$  saturates  the 
Froissart bound. 
In the modern context, the Heisenberg model was updated and generalized in \cite{Nastase:2015ixa}. 
Further work on the Heisenberg model was hindered by the puzzle of how to understand the action (\ref{Heisaction})
from the point of view of QCD? Possible effective actions for QCD are  described in terms of pions that are $SU(2)$-valued.
Skyrme-type actions admit static solitons that can be identified with the nucleons, but do not  have shockwave solutions. 
On the other hand  $SU(2)$-valued generalizations of (\ref{Heisaction}) that have shockwave 
solutions were found not to have static solitons
\cite{Nastase:2005pb}, which presents another puzzle.

 Recently\cite{Cavaglia:2016oda,Bonelli:2018kik,Rosenhaus:2019utc}  it was discovered   that 
 in 1+1 dimensions there are other interesting generalizations
of the DBI scalar action, namely $T\bar T$ deformations of a canonical scalar action with a potential $V$.
 $T\bar T$ deformations have  served as laboratories for investigating  various aspects 
of field theory. See for instance  
\cite{Guica:2017lia,Chen:2018keo,Aharony:2018bad,Cardy:2018sdv,Datta:2018thy,Taylor:2018xcy,Brennan:2019azg,Babaei-Aghbolagh:2020kjg}. 
 
A natural  question that follows from the generalization of Heisenberg's action is:   can such an action be considered 
instead of the action (\ref{Heisaction}) for modeling nucleon scattering, and can it be related to  QCD? After all, in 
\cite{Nastase:2015ixa} we have already found that various generalizations of (\ref{Heisaction}) work as well, and some have 
an interpretation from the point of view of AdS/CFT, in possible gravity duals of QCD. 

With that in mind, we show  in this paper that  certain  $T\bar T$ deformed actions,  indeed admit  
 shockwave solutions which result in the saturation of the Froissart bound, as does the original  model of Heisenberg. 
Moreover, they also have soliton solutions, which upon  boosting allow us to interpret the shockwave as  a soliton collision. 

Our next step is to elevate the 1+1 dimensional abelian deformed theories into a  non-Abelian  one. For that purpose we  make use of 
 the proposal of \cite{Bonelli:2018kik} for a $T\bar T$ deformation of 
the two dimensional WZW model  with a mass term. The deformed  WZW action  is written down 
for $U\in U(N_f)$ with a level $N_c$ where $N_f$ is the number of flavors and $N_c$ the number of colors. 
It was shown in \cite{Frishman:2010zz} that this describes the
 strong coupling effective action of multi-flavor  bosonized  QCD in two dimensions. It was 
 further found  in \cite{Frishman:2010zz} that this action reduces, for the lowest energy 
 configuration, to a sine-Gordon action. In a similar way the deformed WZW action reduces to a $T\bar T$ deformation of  
 the sine-Gordon action. In \cite{HNJS} we found  the soliton solutions of that system and we 
 showed that their masses are the same as  of the un-deformed ones. We argue that these may be used to  describe 
 the physics  of two dimensional nucleons. 

We then uplift this system into a 3+1 dimensional  $T\bar T$-like  deformation 
of the Skyrme action. This action admits   both Skyrme-like solitons
representing the nucleons, and shockwave solutions corresponding to their collisions. We  argue that the latter can be gotten  from 
boosting the former. 
Upon expanding the deformed action as a power series in the deformation parameter, we get that the 
leading order terms are those of the Skyrme model followed by terms with higher power of  derivative 
of the  group element field $U$. These higher order terms are then compared with the corresponding 
terms of the chiral perturbation theory action.

We also briefly discuss the description of the model using the holographic gravity/gauge duality. It turns 
out that as a result of the deformation the holographic model admits a change of the sign of a  brane tension.

The paper is organized as follows. In section 2 we review  Heisenberg's model and its generalizations. In section 3 we consider  
$T\bar T$ deformed scalar actions,  solitons and shockwave solutions, argue that the latter can be understood as 
boosting the former, and find that using the mode  results in a saturation of the Froissart bound. 
In section 4 we use the $T\bar T$ deformation of the WZW action plus a mass term as a deformation 
of the strong coupling limit of multi-flavor QCD in 2 dimensions.   We determine the baryon as a soliton solution 
by transforming the action to that of a deformed 
sine-Gordon mode. We discuss boosting this solution into a shockwave solution. 
In section 5 we consider how to find both Skyrme-like solitons to non-Abelian 3+1 dimensional 
generalizations of the $T\bar T$ deformed action, and shockwave-like solutions, and how to understand the latter from boosting the 
former. In section 6 we consider the possible interpretations of the $T\bar T$ deformed actions in QCD, and in AdS/CFT, 
from the point of view of gravity duals, and in section 7 we conclude and present several open questions. 
In Appendix A we show that the $T\bar T$ deformation of the 
{\em 3+1 dimensional} scalar actions can't be used for the Heisenberg model (unlike the oxidation to 3+1 dimensions of the 
1+1 dimensional actions), and in Appendix B we show that the $T\bar T$ deformations of Abelian and  non-Abelian gauge theories
(Maxwell and Yang-Mills) don't have the needed soliton solutions, so hence cannot be used as a replacement of  the Heisenberg model.

\section{Review of Heisenberg model for high-energy nucleon-nucleon scattering and its generalizations}

In this section we review the model by Heisenberg written in 1952 \cite{Heisenberg1952} to 
describe high-energy nucleon-nucleon scattering, which 
gives a saturation of the Froissart bound \cite{Froissart:1961ux,Lukaszuk:1967zz}, and its modern 
implementation and generalizations in \cite{Nastase:2015ixa}.

Heisenberg starts with the observation that when boosting nucleons to ultra-relativistic speeds, they Lorentz contract, becoming 
first pancake-shaped, until finally they look like delta-function shockwaves in the direction of propagation. Moreover, the pion field 
around them also Lorentz contracts, so finally we can consider that it becomes a delta-function sourced pion field shockwave. 

The process of high-energy nucleon scattering then is described by the collision of two such shockwaves, and as such can be 
described by a (classical)
shockwave solution of an effective action, which was chosen to be (\ref{Heisaction}), for reasons to be explained shortly. 
In the next section, we will see that that in fact the shockwave solution we consider describes the collision of two such 
nucleon-sourced pion shockwaves, and not just one pion shockwave, a fact that was not clear in previous analyses of the model. 

The shockwave solution of (\ref{Heisaction}) is described as a solution depending only on $s=t^2-x^2$, where $x$ is the direction 
of propagation, $\phi=\phi(s)$, with $\phi(s<0)=0$. To constrain the action, Heisenberg imposes that, while $\phi(s)$ must be
continuous, so $\phi(0+)=0$, $(\d_\mu\phi)^2$ must be finite and nonzero on the nontrivial side ($s>0$) and zero on the trivial side, 
the jump in the Lorentz-invariant derivative signalling the existence of the (delta function) nucleon-sourced shock. It is 
found that a canonical scalar with any polynomial potential does not solve it, but the action (\ref{Heisaction}) does, as the solution 
near $s=0$ is 
\be
\phi(s)\simeq l^{-2}\sqrt{s}+...\;,\label{appsol}
\ee
so that 
\be
(\d_\mu \phi)^2\simeq -l^{-4}+...
\ee

On the other hand, in \cite{Nastase:2015ixa} it was found that the condition of jump in derivative, correlated with the existence 
of saturation of the Froissart bound, while it is highly nontrivial, it also admits certain generalizations. The simplest one is to 
an arbitrary potential $V$ inside the square root, 
\be
{\cal L}=l^{-4}\left[1-\sqrt{1+l^4[(\d_\mu\phi)^2+2V(\phi)]}\right]\;,
\ee
another being to several scalars with a sigma model, potential and 2 more functions, 
\be
{\cal L}=l^{-(d+1)}\left[h(\phi^i)-f(\phi)\sqrt{1+l^{d+1}[g_{ij}(\phi^k)(\d_\mu\phi^i)(\d_\mu\phi^j)+ 2V(\phi^i)]}\right].\label{dbimetricv}
\ee

We can also consider an action with an AdS/CFT interpretation, as a D-brane action moving in a curved background, with 
\be
S_{Dd}= l^{-(d+1)}\int d^{d+1} x  e^{-\tilde\phi} \left [ \sqrt{-\det\left( \eta_{\mu\nu}\tilde g(\phi)+ l^{d+1}
\pa_\mu\phi^i\pa_\nu\phi^j g_{ij}(\phi) \right )} -1  \right ]\;,\label{dbraneaction}
\ee
and one can also add some vector mesons (understood as vectors on the D-brane), with Lagrangian
\be
{\cal L}=l^{-4}\left[1-\sqrt{\det(\eta_{ab}+l^4\d_a\phi \d_b\phi+l^2F_{ab})+m^2\phi^2+M_V^2A_a^2}\right].
\ee

The Hamiltonian (energy density) coming from (\ref{Heisaction}) is 
\be
{\cal H}=\pi\dot\phi-{\cal L}=\frac{l^{-4}+(\nabla\phi)^2+m^2\phi^2}{\sqrt{1+l^4[(\d_\mu\phi)^2+m^2\phi^2]}}-l^{-4}\;.\label{Hamden}
\ee

We interpret the classical shockwave solution $\phi(s)$, which, as we will explain in the next section, describes the {\em collision} 
of two pion schockwaves around the high-energy nucleons, as describing in a general direction in $(x,t)$ the pion radiation coming 
from the collision. Indeed, a classical bosonic field is nothing but the quantum field in the case of many bosons in each 
occupied state, so the $\phi(s)$ solution can describe radiation from a collision. Then the Hamiltonian above corresponds to the 
energy radiated in the collision. 

Following Heisenberg \cite{Heisenberg1952}, we can assume that the denominator becomes a non-vanishing constant 
near $s=0$, thus avoiding the unphysical divergence in the energy density near the shock. Fourier transforming in $x$ to momentum 
$k$ the solution near $s=0$ in (\ref{appsol}), we obtain 
\bea
\phi(k,t)&=&l^{-2}\int_0^t dx e^{ikx}\sqrt{t^2-x^2}\cr
&\simeq& l^{-2}\frac{\pi}{2}\frac{|t|}{|k|}(J_1(|k||t|)+i\bf{H}_1(|k||t|))\;,\label{phikt}
\eea
where $J_1$ is a Bessel function and $\bf{H}_1$ is a Struve function. Expanding at large $k$, we obtain 
\be
\phi-l^{-2}i\frac{|t|}{|k|}\propto |t|^{1/2}|k|^{-3/2}e^{-i|k||t|}\;,\label{phiofkandt}
\ee
and dropping the non-oscillatory (non-radiative) piece in $\phi$, we get $\phi(k)\propto k^{-3/2}$. 

However, in reality (and also as we will see in an example in the next section), the shockwave should have a finite thickness of the
order of the Lorentz contracted $1/m$, i.e., $\sqrt{s}_{0m}\equiv \sqrt{s}_{\rm min}=\sqrt{1-v^2}/m$, so at sufficiently large times $t$, 
the momenta $k$ are cut off at $k_{0m}=1/\sqrt{s}_{0m}=\gamma m$, the relativistic mass of the pion. Then the energy of 
radiated pions per unit of spatial momentum, derived from the momentum space Hamiltonian in (\ref{Hamden}) with constant denominator,
is
\be
\frac{dE}{dk}\propto k^2\phi(k)^2\sim \frac{\rm const.}{k}\;,
\ee
for $k\leq k_{0m}$. 

Next we relate to the quantum mechanical description of the classical field, by identifying $k$ with the momentum $k_0$ of a pion, 
and identifying through canonical quantization  the pion field energy $E$ with the radiated energy ${\cal E}$ of the pions, the 
radiated energy per unit of pion energy is 
\be
\frac{d{\cal E}}{dk_0}=\frac{B}{k_0}\;, \;\;\;\;
m\leq k\leq k_{0m}\;,
\ee
integrated to
\be
{\cal E}=B\ln\frac{k_{0m}}{m}= B \ln\gamma. 
\ee
This also leads to a relation for the number of radiated pions per unit of radiated energy, since $dE=k_0dn$ leads to
\be
\frac{dn}{dk_0}=\frac{B}{k_0^2}\;,\;\;\; m\leq k_0\leq k_{0m}\;,
\ee
integrated to
\be
n=\frac{B}{m}\left(1-\frac{m}{k_{0m}}\right).
\ee

Finally, the average emitted energy per pion is
\be
\langle k_0\rangle \equiv\frac{{\cal E}}{n}=m\frac{\ln(k_{0m}/m)}{1-m/k_{0m}}=m\frac{\ln \gamma}{1-\frac{1}{\gamma}}\simeq m\ln \gamma\;,
\ee
which is approximately constant (the dependence on the energy is only logarithmic). On the contrary, for a canonical scalar with 
polynomial potential (and most other canonical scalar actions), 
through a similar calculation, we find at leading order in $\gamma$, $\langle k_0\rangle \propto \gamma$. 

To connect to the saturation of the Froissart bound, we need to consider the fully 4-dimensional form of the action (\ref{Heisaction}). 
For a Mandelstam variable $\tilde s$, the total energy of the colliding nucleons is $\sqrt{\tilde s}$, and we assume that the 
emitted energy is proportional to it, with constant of proportionality given by the wave function overlap. Moreover, since at large
transverse distance $r$ ($=\sqrt{y^2+z^2}$), the wave function is small, so satisfies the free massive KG equation, with solution 
$\phi\sim e^{-mr}$, the wave function overlap will be $e^{-mb}$, where $b$ is the impact parameter of the colliding nucleons. Then 
\be
{\cal E}\sim \sqrt{\tilde s}e^{-mb}.
\ee

The maximum impact parameter, $b_{\rm max}$, arises when we emit a single pion, that is, when the emitted energy ${\cal E}$ 
equals the average per pion emitted energy $\langle k_0\rangle$, so 
\bea
\sqrt{\tilde s}e^{-mb_{\rm max}}&=&\langle k_0\rangle\Rightarrow b_{\rm max}=\frac{1}{m}\ln \frac{\sqrt{\tilde s}}{\langle k_0\rangle}\Rightarrow\cr
\sigma_{\rm tot}&=&\frac{\pi}{m^2}\ln^2\frac{\sqrt{\tilde s}}{\langle k_0\rangle}.
\eea

We see that the saturation of the Froissart bound is equivalent with the (approximate) independence of $\langle k_0\rangle$ of 
the collision energy $\sqrt{\tilde s}$, or in other words, on $\gamma$.

\section{$T\bar T$-like actions as generalizations of the Heisenberg model}

One definition of the $T\bar T$ deformation of a Lagrangian, proposed by Zamolodchikov \cite{Zamolodchikov:2004ce,Smirnov:2016lqw}
is that, for a full quantum theory 
in 1+1 dimensions, 
the variation of the Lagrangian with respect to the deformation parameter equals the determinant of the 
deformed energy-momentum tensor, i.e,
after Wick rotating to Euclidean space, 
\be
\frac{\d {\cal L}}{\d \lambda}=-4[T^\lambda_{zz}T^\lambda_{\bar z\bar z}-(T^\lambda_{z\bar z})^2]\;.\label{equ}
\ee

Note that here all objects are renormalized and UV finite, and on the right-hand side we have an operator regularized by point-splitting. 
We can solve the equation by expansion in a series in $\lambda$ (by treating the fields as numbers), if we give a 
starting point (unperturbed Lagrangian
${\cal L}_0$) and a perturbation parameter $\lambda$ (then by (\ref{equ}), the first order in the series, 
${\cal L}_1$ is given by the unperturbed $\det T_{\mu\nu}$
above). 

Consider as a starting point a real scalar with potential $V$ (still in Euclidean space), 
\be
{\cal L}_0=\frac{1}{2}(\d_\mu\phi)^2+V\;,
\ee
and a perturbation parameter $\lambda$.

Then, at first one obtained a complicated expression, 
with an infinite series of complicated hypergeometric functions (see eq. 6.34 in \cite{Cavaglia:2016oda}), but the series can be summed, 
to obtain a simple 
expression\cite{Bonelli:2018kik},  giving
\bea
{\cal L}(\lambda,X)&=&-\frac{1}{2\lambda}\frac{1-2\lambda V}{1-\lambda V}+\frac{1}{2\lambda}
\sqrt{\frac{(1-2\lambda V)^2}{(1-\lambda V)^2}+2\lambda \frac{X+2V}{1-\lambda V}}\cr
&=&\frac{-(1-2\lambda V)+\sqrt{1+2\bar\lambda (\d_\mu \phi)^2}}{2\bar\lambda}\cr
&\equiv& \tilde V(\phi)+\frac{1}{2\bar\lambda}\sqrt{1+2\bar\lambda(\d_\mu\phi)^2}\;,\label{Lagra}
\eea
where we have used the 
notation $X=(\d_\mu\phi)^2$, as well as defined
\be
\bar \lambda\equiv \lambda(1-\lambda V).
\ee

Note that when $\lambda V\rightarrow 0$, $\tilde V\rightarrow V-1/(2\lambda)$, which 
cancels the a constant coming from the square root, and gives the 
unperturbed potential. 
This can then also be written as in \cite{Rosenhaus:2019utc} (since $4\d \phi \bar \d \phi=(\d_\mu \phi)^2$),
\be
{\cal L}_E= \frac{V}{1-\lambda V}+\frac{-1+\sqrt{1+8\bar \lambda \d \phi \bar \d \phi}}{2\bar \lambda}.\label{LagrangE}
\ee

This is the Lagrangian in Euclidean space. Going back to the Minkowski signature, we obtain 
\be
{\cal L}_M=-\frac{V}{1-\lambda V}+\frac{1-\sqrt{1+2\bar \lambda \d_\mu \phi \d^\mu \phi}}{2\bar \lambda}\;,\label{Lagrang}
\ee
with $\d_\mu \phi \d^\mu \phi =-(\dot\phi)^2+(\phi')^2\equiv X$, as usual. 

We now examine whether these Lagrangians (for various possible potentials $V(\phi)$) can be used for 
generalizations of the Heisenberg model
(other than the previously considered ones reviewed in the previous section). 

\subsection{Review of soliton solutions}

In this subsection we review soliton solutions described in a companion paper \cite{HNJS} (after the paper was posted, 
we became aware of \cite{Conti:2018jho,Conti:2018tca}, which have some overlap with ours, though here we present the 
solutions in the form we obtain)

We have considered $T\bar T$ and Heisenberg deformations of a canonical scalar with a potential $V$. 

\subsubsection{Heisenberg deformations}

Starting with the Heisenberg case, the deformed Lagrangian  in 1+1 dimensional (Minkowski) space is 
\be
{\cal L}_M=1-\sqrt{1+(\d_\mu\phi)^2+2V}\;,
\ee
and, by use of a modified virial theorem, we have found that the generic static solution is written implicitly as 
\be
x-x_0=\int_{\phi(x_0)}^{\phi(x)} \frac{d\phi}{\sqrt{(1+2V)\left[\frac{1+2V}{C^2}-1\right]}}\;,
\ee
where $C$ is an arbitrary constant.

By comparing with the canonical case, we have found that we need to put $C=1$ to find a modified soliton solution, so that 
the implicit soliton solution is 
\be
x-x_0=\int_{\phi(x_0)}^{\phi(x)}\frac{d\phi}{\sqrt{2V(1+2V)}}\;.
\ee

For the pure DBI scalar case, with $V(\phi)=0$, the solution is 
\be
x-x_0=\int_{\phi(x_0)}^{\phi(x)}\frac{2d \phi}{\sqrt{\frac{1}{C^2}-1}}=K(\phi(x)-\phi(x_0))\;,
\ee
and by gluing two such solutions we find the solution to the Poisson equation in one dimension, 
\be
\phi(x)=\phi(x_0)+K^{-1}|x-x_0|.
\ee

We note that, if we redefine $V$ such that $V(0)=0$, this solution is valid also for the case of nonzero $V(\phi)$, just that only 
near $x=x_0$. After an infinite, or rather very large, boost, with $K^{-1}\gamma\equiv \tilde K$, we have the 
propagating shockwave solution
\be
\phi(x^-)=\tilde K|x^-|.
\ee

But more relevant, perhaps, is the case of the pure DBI action {\em in 3+1 dimensions}, 
in which case the conservation equation becomes (using the same formalism as in 1+1 dimensions, but
for the one-dimensional Lagrangian in the radial direction $r=x$, ${\cal L}(r)=4\pi r^2[1-\sqrt{1+\phi'^2}]$)
\be
\frac{d}{dx}\left(x^2\frac{1+2V}{\sqrt{1+2V+\phi'^2}}\right)=-2x\sqrt{1+2V+\phi'^2}.
\ee

Here it is harder to write the general solution, but one can find the specific solution ("catenoid", see later sections)
with a delta function source at $r=0$, solution 
of $\vec{\nabla}\cdot \vec{C}=q\delta^3(r)$, with $\vec{C}=\d {\cal L}/\d(\vec{\nabla}\phi)$, which is 
\be
\phi'^2=\frac{q^2}{x^4-q^2}\Rightarrow \phi (r)=q\int_r^\infty \frac{dx}{\sqrt{x^4-q^2}}.
\ee

\subsubsection{$T\bar T$ deformations}

In the case of $T\bar T$ deformations of a canonical scalar with a potential $V$, we can find the implicit general static
solution for $\phi(x)$ as 
\be
x-x_0=\int_{\phi(x_0)}^{\phi(x)}\frac{d\phi \sqrt{2\lambda (1-\lambda V)}}{\sqrt{\frac{1}{[2\lambda (1-\lambda V)(C-\tilde V)]^2}-1}}\;,
\ee
where again $C$ is an arbitrary integration constant. 

The relevant case for a soliton, that arises as a deformation of the soliton for the canonical scalar with potential $V$ is 
obtained for $C=0$, as 
\be
x-x_0=\int_{\phi(x_0)}^{\phi(x)} \frac{d\phi|1-2\lambda V|}{\sqrt{2V}}.
\ee

In particular, for the sine-Gordon potential,
\be
V_{sG}= -\frac{\mu^2}{\beta}\left[\cos(\beta \phi) -1 \right]\;,\label{sG}
\ee
we find the deformed sine-Gordon soliton, defined implicitly as 
\be
\pm \mu\sqrt{\b}(x-x_0)=\left.\left[\ln\left(\tan\frac{\b\phi}{4}\right)+4\frac{\lambda \mu^2}{\b}
\cos\frac{\b\phi}{2}\right]\right|_{\phi(x_0)}^{\phi(x)}\;,
\ee
and for the Higgs-type potential, 
\be
V(\phi)=\a(\phi^2-a^2)^2\;,
\ee
we find the deformed kink soliton, defined implicitly as 
\be
x-x_0=\pm \frac{1}{a\sqrt{2\a}}\left[\tanh^{-1} \frac{\phi}{a}-4\lambda \a a^3 \left|\frac{\phi^3}{3a^3}-\frac{\phi}{a}\right|\right]\;.
\ee

One again finds the same approximate static solution, near a point $x=x_0$, the solution to the Poisson equation in one 
dimension, 
\be
\phi(x)=\phi(x_0)+K^{-1}|x-x_0|\;,
\ee
and as before, after an infinite boost goes over to the same 
\be
\phi(x^-)=\tilde K |x^-|.
\ee

We should also consider the case of soliton solutions of the {\em 3+1 dimensional extension} of the $T\bar T$ Lagrangian, 
for a spherically symmetric solution, when we can write a one-dimensional radial Lagrangian for the radius $r=x$ via
${\cal L}(x)\rightarrow 4\pi r^2 {\cal L}(r)$. Then, doing exactly the same steps as before, we obtain the conservation equation 
\be
\frac{d}{dx}\left[\frac{x^2}{2\lambda(1-\lambda V)\sqrt{1+2\lambda (1-\lambda V)\phi'^2}}+x^2\tilde V\right]=-2x\frac{\sqrt{1+2
\lambda (1-\lambda V)\phi'^2}}{2\lambda (1-\lambda V)}\;,
\ee
or, formally integrating to an integro-differential equation for $\phi$ and $\phi'$, 
\be
\frac{x^2}{2\lambda(1-\lambda V)\sqrt{1+2\lambda (1-\lambda V)\phi'^2}}+x^2\tilde V=C-\int dx \; 2x\frac{\sqrt{1+2
\lambda (1-\lambda V)\phi'^2}}{2\lambda (1-\lambda V)}.
\ee

Alternatively, we could obtain an integro-differential equation by defining $\vec{E}\equiv \vec{\nabla}\phi$ and 
\be
\vec{D}\equiv \frac{\d {\cal L}}{\d \vec{E}}\;,\;\;\; \rho(\vec{r})\equiv \frac{\d {\cal L}}{\d \phi(\vec{r})}\;,
\ee
so that the equation of motion is $\vec{\nabla}\cdot \vec{D}(\vec{r})=\rho(\vec{r})$, like for electromagnetism in a medium. 
Since for $\rho(\vec{r})=\delta^3(\vec{r}-\vec{r}')$ we have as solution for $\vec{D}$ the Green's function 
\be
\vec{D}_0(\vec{r}-\vec{r}')=\frac{\widehat{\vec{r}-\vec{r}'}}{4\pi |\vec{r}-\vec{r}'|^2}\;,
\ee
the solution for $\vec{D}$ is 
\be
\vec{D}(\vec{r})=\int d^3\vec{r}'\rho(\vec{r}')D_0(\vec{r}-\vec{r}')\;,
\ee
so we obtain the integro-differential equation for $\phi$ and $\vec{\nabla}\phi$ ($\phi'$ on the radial ansatz)
\be
\frac{\d {\cal L}}{\d \vec{\nabla}\phi(\vec{r})}=\int d^3\vec{r}'\frac{\d {\cal L}}{\d \phi (\vec{r}')}\frac{\widehat{\vec{r}-\vec{r}'}}{4\pi |
\vec{r}-\vec{r}'|^2}.
\ee

However, this is more complicated than the integro-differential equation obtained before, 
so we don't gain anything (in the pure DBI case, on the right-hand side one had 
only $qD_0(\vec{r})$, allowing us to completely solve the equation).

\subsection{The shockwave solution}

In the previous subsection we have considered static solutions $K|x-x_0|$ near $x=x_0$, infinitely boosted to 
solutions $\tilde K|x^-|$, representing shockwaves propagating at the speed of light. 

However, there are other shockwave solutions, namely ones described by a function of only $s=t^2-x^2=x^+x^-$ ($x^\pm =
t\pm x$), and not independently on $x^+,x^-$. In this case, in \cite{HNJS} we have found that both for the Heisenberg and 
$T\bar T$ deformations, we find the same perturbative shockwave solution as Heisenberg: near $s=0$, but $s>0$, the solution is 
\be\label{shwa}
\phi(s)=l^{-2}\sqrt{s}+{\cal O}(s^{3/2})\;,
\ee
while $\phi(s<0)=0$.

We have also found that the solution remains true if we change the square root in the action 
with another rational power less than 1, $p/q<1$. 

In the next subsection, we turn to the interpretation of this solution.

\subsection{The shockwave solution as collision of {\em two} ultrarelativistic waves representing high energy nucleons}

In the previous section, as well as in \cite{Nastase:2015ixa}, the 1+1 dimensional solution was described as 
a shockwave, and it is, in the sense
of being a function of $s$ that has discontinuous derivatives at $s=0$. However, from a physical point of view, 
we will see now that it actually describes 
{\em two} shockwaves colliding, which means it really corresponds to the ultra-relativistic limit of the field of 
colliding moving particles sourcing it, identified with 
the high energy nucleons. 

To see this, we first calculate the energy momentum tensor.
Including a nontrivial metric in Heisenberg's Lagrangian for the pion, of the type of a DBI scalar with a mass term in the square root, 
\be
{\cal L}=l^{-4}\sqrt{-g}\left[1-\sqrt{1+l^4(\d_\mu\phi\d_\nu\phi g^{\mu\nu}+m^2\phi^2)}\right]\;,
\ee
and varying it, gives the Belinfante tensor
\bea
T_{\mu\nu}&=&\frac{-2}{\sqrt{-g}}\frac{\delta S}{\delta g^{\mu\nu}}=\frac{\d_\mu\phi\d_\nu\phi}{\sqrt{1+l^4(\d_\mu\phi\d_\nu\phi g^{\mu\nu}+m^2\phi^2)}}\cr
&& +g_{\mu\nu}l^{_4}\left[1-\sqrt{1+l^4(\d_\mu\phi\d_\nu\phi g^{\mu\nu}+m^2\phi^2)}\right].
\eea

But, as we found in \cite{Nastase:2015ixa}, 
for $s=t^2-x^2\rightarrow 0$, on the solution, the square root Lagrangian diverges 
(and we argued that it must be regularized somehow, since divergencies
will lead to quantum fluctuations). But, on the solution, with $x^\pm= x\pm t$, so $s=-x^+x^-$,
\be
\frac{\d\phi}{\d x^+}=\frac{d\phi}{ds}\frac{\d s}{\d x^+}=-x^-\frac{d\phi}{ds}\;\;\; \frac{\d \phi}{\d x^-}=-x^-\frac{d\phi}{ds}.
\ee

Then, on the solution, near $s=0$, the term with $g_{\mu\nu}l^{-4}$ is finite, 
the term with $g_{\mu\nu}$ times the square root goes to zero, and 
the leading term diverges, as 1 over the square root, so 
\bea
T_{++}\simeq -\frac{\d_+ \phi\d_+\phi}{\sqrt{1+l^4(\d_\mu\phi\d_\nu\phi g^{\mu\nu}
+m^2\phi^2)}}\propto \left(\frac{d\phi}{\d s}\right)^2 (x^-)^2\;,\cr
T_{--}\simeq -\frac{\d_- \phi\d_-\phi}{\sqrt{1+l^4(\d_\mu\phi\d_\nu\phi g^{\mu\nu}+m^2\phi^2)}}\propto \left(\frac{d\phi}{\d s}\right)^2 (x^+)^2
\;,\cr
T_{+-}\simeq -\frac{\d_+ \phi\d_-\phi}{\sqrt{1+l^4(\d_\mu\phi\d_\nu\phi g^{\mu\nu}+m^2\phi^2)}}\propto
 \left(\frac{d\phi}{\d s}\right)^2 s\;.\label{enmomten}
\eea

The 1 over the square root diverges, but must be regularized to a finite, yet large value. But then we note that, near $s=0$ (which means $x^+=0$ 
{\em or} $x^-=0$), we have $T_{+-}\rightarrow 0$, and, for $x^+=0$,  $T_{--}\rightarrow 0$, thus only $T_{++}$ is nonzero, and is a regularized divergence, 
{\em and} for $x^-=0$, $T_{++}\rightarrow 0$, thus only $T_{--}$ is nonzero, and is a regularized divergence. 

That means that the shockwave solution of Heisenberg represents {\em two} colliding shockwaves (instead of one), 
one at $x^+=0$ (going in the negative $x$ direction) with 
only $T_{++}$ and diverging (like the A-S shockwave in gravity), and one at $x^-=0$ (going in the positive $x$ direction), with only $T_{--}$ and diverging. 

This is a solution in 1+1 dimensions, so there is no dependence on the 2 transverse directions, and the collision of the two shockwaves happens at $x=t=0$. 

If we view it as a limit $r\rightarrow 0$ of a solution $\phi(s,r)$, then implicitly we assume that the collision happens at $x=r=0$, i.e., head-on collision (no 
impact parameter $b$). Yet it also means that the $\phi(s\simeq 0, r\neq 0)$  represents the wave function {\em overlap} of the two nucleons, as indeed 
Heisenberg claims in his paper, and this overlap also goes like $e^{-m_\pi r}$ at large $r$, as he says also. 

Moreover, as observed both by him implicitly and by us \cite{Nastase:2015ixa} implicitly, 
this $\phi(s\simeq 0, r\simeq 0)\simeq \phi(s\simeq 0)$ is consistent with large spatial momenta $k$, if we consider 
$x$ small and $r$ small, which one did in order to calculate the emitted spectrum (and moreover, we need large $t$ in order for it to be the emission spectrum 
at large time after the collision, i.e., emitted radiation). All of these were conditions imposed on the radiation derived from the classical solution, 
so their presence is consistent. 

But we can view the shockwave solution also as part of a solution $\phi(x^+,x^-,r)$, at $r\rightarrow 0$, but {\em only}
at $x-=0$, $x^+<0$, which means $t<0$ (before 
the collision), but $x<0$, so near one of the shockwaves, before the other one arrives. In that case, we can assume the other one (the one 
going in the negative $x$ direction) can be situated at some nonzero impact parameter, $r=b\neq 0$. That is why the solution can be 
also thought of as {\em representing the collision at some impact parameter $b$}. 

This explains why the solution represents collision of 2 waves, and why we can get the maximum 
impact parameter $b_{\rm max}$ by using it. 

\subsection{On boosting a soliton into a shockwave}

Before continuing with the analysis of the cross section for high energy scattering, an issue to consider is: how to 
understand the relation between static soliton solutions and shockwave solutions for the same action. One would expect that by 
infinitely boosting the soliton solution, one would get the shockwave, or some limit of it. However, as we just saw, the shockwave 
solution actually shows the {\em collision} of two waves (one at $x^+=0$, another at $x^-=0$), so that is not what one wants. 
In fact, we have seen that by boosting a soliton solution we (rather generically) obtain the solution $\phi(x^-)=\tilde K |x^-|$. 

Rather, if we are to compare with the shockwave solution, we should try to compare the energy-momentum tensors, more specifically 
the one coming out of a single wave, say $T_{--}$. Considering the Heisenberg model with a mass term only (neglecting 
possible higher orders in $\phi$ in the potential $V$), which is the only thing that matters near the shock at $s=0$, the near shock solution 
is 
\be
\phi\simeq \frac{\sqrt{s}}{l^2}(1+a \; s\; m^2+{\cal O}(s^2))\;,
\ee
where $a$ is an arbitrary constant. Then the square root defining the Lagrangian, and the Lorentz-invariant 
denominator of the Hamiltonian and of $T_{\mu\nu}$, is 
\be
\sqrt{1+l^4(g^{\mu\nu}\d_\mu \phi \d_\nu\phi +m^2\phi^2)}\simeq m \sqrt{s}\sqrt{1-6a}\;,
\ee
so it diverges as $s\rightarrow 0$. We also note the condition $a\leq 1/6$ necessary for reality of the Lagrangian. Then more precisely, 
we have
\be
T_{+-}\simeq \frac{1}{4\sqrt{1-6a}}\frac{l^{-4}}{m\sqrt{s}}\;,\;\;
T_{--}\simeq \frac{1}{4\sqrt{1-6a}}\frac{l^{-4}}{m\sqrt{s}} \frac{x^+}{x^-}\;\;\;
T_{++}\simeq \frac{1}{4\sqrt{1-6a}}\frac{l^{-4}}{m\sqrt{s}}\frac{x^-}{x^+}.
\ee

Then, if $x^+$ is finite but $x^-\rightarrow 0$, $T_{++}\rightarrow 0$, $T_{+-}\propto 1/\sqrt{x^-}$, and $T_{--}\propto 1/(x^-)^{3/2}$. 
But that means that at the shock $T_{+-}\rightarrow\infty$ also, whereas it should really not diverge, 
at least in the case of a single wave. So, 
it is natural to assume that the square root should somehow be cut off  neat $s=0$ such that it stays finite (in fact, this was 
implicitly assumed in Heisenberg's case; for the Hamiltonian density $H$, but it is the same denominator). In that case we would have 
$T_{++}\rightarrow 0$, $T_{+-}$ finite, $T_{--}\propto 1/x^-$. 

But consider the generic soliton solution of $\phi(x)=\tilde K|x-x_0|$ near $x_0$, which was infinitely boosted to $\phi(x)=\tilde K\gamma 
|x^-|$ (for $\gamma\rightarrow \infty$), 
near $x^-=0$. Moreover, consider the Lorentz invariant $l^4g^{\mu\nu}\d_\mu \phi \d_\nu \phi$ which can be calculated on the 
static solution as $l^4 \tilde K^2$. For the infinite boost to make sense, we need that $\tilde K\rightarrow 0$, $\tilde K\gamma$ finite, 
in which case the invariant is $l^4\tilde K^2\rightarrow 0$. Then the square root on the soliton solution is 
\be
\sqrt{1+l^4m^2\tilde K ^2 \gamma^2 (x^-)^2}\simeq (l^2m\tilde K\gamma)|x^-|\;,
\ee
where we have assumed that the second term is larger than the 1 (we can have $l^2\tilde K \gamma $ large enough  for that). Then 
\be
T_{--}\simeq \frac{\d_-\phi \d_-\phi}{\sqrt{1+l^4(g^{\mu\nu}\d_\mu \phi \d_\nu \phi+m^2\phi^2)}}\simeq \frac{\tilde K \gamma l^{-2}}{m|x^-|}
\ee
has the same behavior as that expected from the shock near one wave (with the cut-off in the square root).

It would seem like the total energy, $\int dx^- T_{--}$, has a log divergence at $x^-=0$.
However, note that for this correct boosted soliton solution, there is a (very small) $x^-$ at which $T_{--}$ remains finite.
More precisely, since 
\be
T_{--}=\frac{\tilde K \gamma l^{-2}}{m\sqrt{(x^-)^2+\frac{1}{l^4m^2\tilde K^2\gamma^2}}}\;,
\ee
we have that 
\be
E\sim \int dx^- T_{--}= \frac{\tilde K \gamma l^{-2}}{m}\sinh^{-1} (x^-l^2m \tilde K \gamma)\;,
\ee
which doesn't have a divergence at $x^-=0$, and the (log) divergence at large $x^-$ is spurious also, because the original boosted 
solution was valid only near $x=x_0$, or near $x^-=0$, when boosted. We also note the correct relativistic relation $E\propto \gamma$.

\subsection{The cross section and saturation of the Froissart bound}

Now that we saw that the (1+1-dimensional)
$T\bar T$ action gives the same shockwave near $s=0$, and moreover the shockwave solution 
represents a collision of two ultrarelativistic waves representing the 
high energy nucleons, we can trivially generalize the action to 
3+1 dimensions, by making $(\d_\mu\phi)^2$ 3+1-dimensional in the Lagrangian, as well as replacing $d^2x$ with $d^4x$,
and use it as Heisenberg did, for the high-energy nucleon-nucleon collision. 

We want to see if the collision cross section can still saturate the Froissart bound, if we use this 3+1 dimensional generalization of the $T\bar T$ action
instead of the action used by Heisenberg (which, in any case, was just a guess, as the {\em simplest}, but not 
necessarily most useful or correct; one that gave him the needed properties of the shockwave for 
high energy scattering of nucleons).

In the Heisenberg analysis reviewed in section 2, which we made more precise in \cite{Nastase:2015ixa}, the next step in getting 
to the cross section of Froissart, is 
to calculate the radiated energy per unit of $k$, so first we need the Hamiltonian. At this point, we 
are still in 1+1 dimensions, so $(\d_\mu\phi)^2=-\dot\phi^2+\phi'^2$. Then 
\bea
{\cal H}&=& \pi \dot\phi-{\cal L}=\frac{\d {\cal L}}{\d \dot \phi}\dot\phi-{\cal L}\cr
&=& -\frac{1}{2\bar\lambda}\frac{\d}{\d\dot\phi}\sqrt{1+8\bar \lambda(-\dot\phi^2+\phi'^2)}\dot \phi
+\frac{1}{2\bar\lambda}\sqrt{1+8\bar \lambda(-\dot\phi^2+\phi'^2)}+\tilde V(\phi)\cr
&=&\frac{1}{2\bar \lambda}\frac{1+8\bar \lambda\phi'^2}{\sqrt{1+8\bar \lambda(-\dot\phi^2+\phi'^2)}}
+\tilde V(\phi).
\eea

But the energy is 
\be
E=\int dx {\cal H}(x)=\int dk {\cal H}(k)\;,
\ee
where now the Hamiltonian density is
\be
{\cal H}(k)=\frac{1}{2\bar\lambda}\frac{1+8\bar \lambda k^2\phi^2(k)}
{\sqrt{1+8\bar \lambda(-\dot\phi^2+k^2\phi^2)}}-\tilde V(\phi(k)).
\ee

But, like in the case of \cite{Nastase:2015ixa} (and of Heisenberg), we see that the first term has the 
square root in the denominator, and on the solution near $s=0$, the square root vanishes (since 
$V$ vanishes there, and then the rest is the same square root as for Heisenberg). But, as Heisenberg implicitly 
noted (and we explained in detail), this is unphysical, and we expect that some quantum fluctuations or corrections 
make the square root a small, but nonvanishing constant near $s=0$. 

Then, neglecting the subleading term $\tilde V(\phi(k))$, we obtain 
\be
{\cal H}(k)\simeq \frac{k^2\phi^2(k)}{\rm const.}.
\ee

But, since the same $\phi(x,t)\simeq l^{-2}\sqrt{s}$ solution is valid near $s=0$ here, as in 
the case of  \cite{Nastase:2015ixa}, we have the same Fourier transform $\phi(k,t)$ at large $k$, 
eqs. (2.14) and (2.15) there. There is a non-oscillatory piece, which can be dropped as there, since we 
are focusing in on the radiative solution. Then $\phi(k)\sim |k|^{-3/2}$, so we have
\be
\frac{dE}{dk}={\cal H}(k)\sim\frac{\rm const.}{k}\;,
\ee
as before. This then leads, as we saw in \cite{Nastase:2015ixa} and in the previous section, after several steps, to 
the average per pion energy being approximately a constant, 
\be
\langle k_0\rangle\simeq m \ln \gamma\;,
\ee
where $\gamma$ is the Lorentz factor for the collision. 

Only now do we need to consider a fully 3+1 dimensional version of the $T\bar T$ action. 
More precisely, we need to use the 
small field value for it, which is just the canonical free massive scalar action. 
Then, the solution at small field, in terms of transverse radius $r=\sqrt{y^2+z^2}$, is 
\be
\phi\sim e^{-mr}\;,
\ee
and then the same argument as Heisenberg's follows: the radiated energy ${\cal E}$ is proportional 
to the total collision energy $\sqrt{\tilde s}$, and to the overlap of wave functions, $\sim e^{-mb}$, 
where $b$ is the impact parameter; and at maximum $b$, we radiate only $\langle k_0\rangle$, 
so $\sqrt{\tilde s}e^{-mb_{\rm max}}=\langle k_0\rangle$, from which we get the cross 
section $\sigma_{\rm tot}=\pi b^2_{\rm max}=\frac{\pi}{m^2}\ln^2\frac{\sqrt{\tilde s}}
{\langle k_0\rangle}$.


\section{From Abelian to non-Abelian  Heisenberg and $T\bar T$ deformed models}

Next,  we would like to consider non-Abelian versions of the Abelian actions considered in the Heisenberg 
type model.  In this section we will consider still 1+1 dimensional models, and in the next one we will 
consider the full 3+1 dimensional models needed. 

For the generalization of the abelian field to non-abelian ones, we can follow two approaches:

(a) Straightforwardly utilizing a $ U\in U(2)$ group element instead of the single real scalar field $\phi$. 

(b) Using the proposal of \cite{Bonelli:2018kik} for a $T\bar T$ deformation of 
the two dimensional WZW model, deformed also by the potential (mass) term.
In this paper, the deformation of the WZW model was found, and the 
kinetic term is replaced by a square root term, while 
the WZW term is unchanged by the $T\bar T$ deformation, being topological.

We first consider the  case (a),  of 
Heisenberg and $T\bar T$ deformations, 
with a potential of 
\be
V= -m^2 \left[\cos\left(\sqrt{\frac{4\pi}{N_c}}\phi(t,x)\right)-1\right]  \;, 
\ee
where $N_c$ is the number of colors.
To generalize to a non-abelian case we make the following replacements
\bea 
\pa_\mu\phi(t,x)\pa^\mu\phi(t,x) \qquad &\rightarrow & \qquad \frac{N_c}{4\pi} \Tr\left[\pa_\mu U(t,x)\pa^\mu U^\dagger(t,x)\right]\CR
\cos\left(\sqrt{\frac{4\pi}{N_c}}\phi\right)-1 \qquad &\rightarrow & \qquad \half \Tr\left[U+ U^\dagger -2\right].\CR 
\eea

The non-Abelian Lagrangian density corresponding to the Abelian $T\bar T$ deformation (\ref{Lagrang})
(Minkowskian) then reads
\bea\label{naaction2} 
{\cal L}_M &=& \frac{\left [ 1 - \sqrt{1+ \frac{N_c}{2\pi}\lambda\left(1+\frac{\lambda m^2}{2} \Tr\left[U+ U^\dagger -2\right]\right) 
\Tr\left[\pa_\mu U(t,x)\pa^\mu U^\dagger(t,x)\right]}\right ]  }{2\lambda(1+\frac{\lambda m^2}{2} \Tr\left[U+ U^\dagger -2\right])}\cr
&&+ \frac{m^2 \Tr\left[U+ U^\dagger -2\right]}{2(1+\frac{\lambda m^2}{2} \Tr\left[U+ U^\dagger -2\right])}.
\eea

Note that whereas the kinetic part of the action (\ref{naaction2}) is invariant under 
$U_L(2)\times U_R(2)$ symmetries associated with the transformations
$U\rightarrow A_L U$,  $U\rightarrow  UA_R$,
where $A_L$ and $A_R$ are constant $U(2)$ matrices, the potential term and hence the whole action is 
only invariant under the diagonal $U(2)$ transformation (with $A_L=A_R^\dagger=A$)
\be
U\rightarrow A U A^\dagger.
\ee

In particular the  diagonal $U(1)_D$ symmetry should be associated with the baryon number, according to the general theory
of QCD.
The corresponding Noether currents, for $U(2)$ generators $T^A=(\frac{\sigma^a}{2},\one)$ (where $\sigma^a$ are the Pauli matrices) 
are given by
\be
j_\mu^A=-i\frac{N_c}{4\pi}\frac{\Tr\left[\d_\mu U T^A U^\dagger -U^\dagger T^A\d_\mu U^\dagger\right]}{
\sqrt{1+ \frac{N_c}{2\pi}\lambda\left(1+\frac{\lambda m^2}{2} \Tr\left[U+ U^\dagger -2\right]\right) 
\Tr\left[\pa_\mu U(t,x)\pa^\mu U^\dagger(t,x)\right]}}.
\ee

Note that in the Skyrme model, the presence of the extra Skyrme term means that the Noether current has an extra 
contribution equal to the topological current (for the topological number associated with the hedgehog ansatz). On a static ansatz
$\d_0 U=0$, the above zero component of the
Noether current, as the corresponding term coming from the kinetic term in the Skyrme model, vanishes, meaning that
the Noether charge equals to just the topological charge, and is associated with baryon number. Here, however, on the 
static ansatz we just get a zero Noether charge.

This suggests that perhaps it is better to use the case (b), for the proposal in \cite{Bonelli:2018kik}, of deforming the WZW model, 
which will have a nonzero Noether current, thus a nonzero Noether charge, that will turn out to be equal to the topological charge, and 
thus can be equated with the baryon number, for the dimensional reduction of an effective model for QCD ($QCD_2$). 
The model in \cite{Bonelli:2018kik} will also be deformed by a potential (mass) term.

In conclusion, in this section we will analyze the deformations of  
(i)  the  WZW model (ii) the  massive WZW model and 
(iii) baryonic sector of bosonized $QCD_2$ in the strong coupling limit.

\subsection{The deformation of the WZW theory}

It is well known that in 1+1 dimensions the theory of a free massless real scalar  described by the action
\be
S= \frac{1}{8\pi}\int d^2 x \;\pa_\mu\phi \pa^\mu\phi = \frac{1}{4\pi}\int d^2 z \pa\phi\bar\pa\phi\;,
\ee
where the second form of the action is taken in complex plane,
is invariant under transformations with parameters which are holomorphic and 
anti-holomorphic with the corresponding $U(1)$ affine Lie algebra conserved 
currents
\be
J=\pa \phi\;,\qquad \bar J =\bar \pa \phi\;, \qquad \pa\bar J=\bar\pa J=0\;,
\ee
and energy momentum tensor
\be
T=-\frac12\pa \phi\pa \phi\;,\qquad  \bar T=-\frac12\bar\pa \phi\bar\pa \phi\;,\qquad \pa\bar T=\bar\pa T=0.
\ee

The $U(N)\times U(N)$ non-abelian generalization of this system is the well-known WZW action given by
\be
S=S_{\sigma}+S_{WZW}=\frac{1}{2}\int d^2x \frac{k}{4\pi}\Tr[\d_\mu U 
\d^\mu U^\dagger]+\frac{k}{12\pi}\int_{B^3}   [\Tr U^{-1}d U]^3\;,
\ee
where $k$ is the level of the Kac Moody algebras associated with the 
holomorphic (anti-holomorphic)  conserved  currents $ \bar J$ ($\bar J$) given by
\be
J= \frac{k}{4\pi}\pa U U^{-1}\;, \qquad \bar J \frac{k}{4\pi}  U^{-1}\bar \pa U\;,\qquad \pa \bar J = \bar\pa J=0.
\ee

The corresponding energy momentum tensor is given by the Sugawara construction
\be
T= \frac{1}{k} Tr[ J J]\;, \qquad \bar T= \frac{1}{k} Tr[ \bar J \bar J]\;,\qquad 
\pa \bar T = \bar\pa T= T_{z\bar z}=0.
\ee

The first term in the action, the sigma term, follows from a map
\be\label{mapU}
\pa_\mu\phi(t,x)\pa^\mu\phi(t,x) \qquad \rightarrow  \qquad \frac{k}{4\pi} \Tr\left[\pa_\mu U(t,x)\pa^\mu U^\dagger(t,x)\right].
\ee

The $T\bar T$ deformation of the Euclidean  WZW action was worked out 
in \cite{Bonelli:2018kik}. It takes the following form  
\be
S=\int d^2x \left[-\frac{1}{2\lambda}+\frac{1}{2\lambda}\sqrt{1+\frac{N_c}{2\pi}\lambda X 
+\frac{N_c}{2\pi}\frac{N_c}{4\pi}\lambda^2\tilde X^2}\right]
+\frac{i}{4\pi}\int_{B^3} \frac{N_c}{4\pi} [\Tr U^{-1}d U]^3\;,
\ee
 where 
\bea
X&\equiv &\Tr [\d_\mu U \d^\mu U^\dagger]=-\Tr[L_\mu L^\mu]\cr
L_\mu&\equiv& U^{-1}\d_\mu U\cr
\tilde X^2 &\equiv &  \epsilon^{\mu\rho}\epsilon^{\nu\sigma}\Tr[L_\mu L_\nu]\Tr[L_\rho L_\sigma].
\eea

Notice that  the WZW term is  undeformed. We have also identified $k=N_c$, we will see in the following subsections why.
When  expanding in small $\lambda$ the action takes the form
\bea
S&=&\frac{N_c}{4\pi}\int d^2x \left[\frac{X}{2}+\frac{\lambda N_c}{2\pi} \left(\frac{(\tilde X)^2}{4}-\frac{X^2}{8}\right)+\frac{1}{16} 
\left(\frac{\lambda N_c}{2\pi}\right)^2
   \left(X^3-2 X (\tilde X)^2\right)+{\cal O}\left(\lambda^3\right)\right]\cr 
   &&+ S_{WZ}.
\eea

It is easy to check that both $X$ and $\tilde X_{ij}^2$ are invariant under the left and right $U(N)$ transformations 
\be
 U\rightarrow A_L U\;,\qquad A_L\in U_L(N)\;,
 \qquad U\rightarrow  U A_R\;,\qquad A_R\in U_R(N)\;,
\ee
and hence these are symmetry transformations of the full deformed action.
The corresponding currents $J^\mu_L$ and $J^\mu_R$ are given by 
 \be
{J_\mu^A}_{(L,R)} =i\frac{N_c}{4\pi}\left(\frac{\Tr\left[-\d_\mu U T^A U^\dagger +U^\dagger T^A\d_\mu U^\dagger\right]}{
\sqrt{1+ \lambda\frac{k}{2\pi} 
\Tr\left[\pa_\mu U(t,x)\pa^\mu U^\dagger(t,x) \right]}} 
\pm\epsilon_{\mu\nu}
\Tr\left[\d^\nu U T^A U^\dagger -U^\dagger T^A\d^\nu U^\dagger\right]\right),
\ee  
where $T^A$ are the generators of the $U(N)$ group. The left and right currents correspond to the different signs  $\pm$,
{\em are not conserved holomorphicaly  and anti-holomorphically}, but rather in  the ordinary covariant form 
\be
\pa_\mu J^\mu_L= 0\;, \qquad  \pa_\mu J^\mu_R= 0.
\ee

Similarly  for the energy momentum tensor, and furthermore for the deformed theory  we have $T_{z\bar z}\neq 0$.

\subsection{WZW with a ``mass term"}

So far we discussed the deformation of the WZW theory, the non-abelian generalization 
of single free massless scalar field. Next we would like to address the same procedure 
for an interacting scalar field, namely, a scalar field with a potential, since this is closer to the model we seek.  
More specifically we consider a sine-Gordon potential given by
\be
V= -m^2 \left[\cos\left(\beta\phi(t,x)\right)-1\right].  \ 
\ee

To map the abelian theory  to a non-abelian  we  generalize the map of (\ref{mapU}) with  the following replacements
\bea 
\pa_\mu\phi(t,x)\pa^\mu\phi(t,x) \qquad &\rightarrow & \qquad \frac{k}{4\pi} \Tr\left[\pa_\mu U(t,x)\pa^\mu U^\dagger(t,x)\right]\CR
\cos\left(\beta\phi\right)-1 \qquad &\rightarrow & \qquad \half \Tr\left[U+ U^\dagger -2\right].\CR 
\eea

The  {\em Minkowski space } action that corresponds to the $T\bar T$ deformation of this system is given by

\bea
S_M &=& \int d^2x\left\{\frac{\left [ 1 - \sqrt{1+ \frac{k}{2\pi}\lambda\left(1+\frac{\lambda m^2}{2} \Tr\left[U+ U^\dagger -2\right]\right) 
\left(\Tr\left[\pa_\mu U(t,x)\pa^\mu U^\dagger(t,x)\right]+\frac{k}{4\pi}\lambda\tilde X^2\right)}
\right ]  }{2\lambda(1+\frac{\lambda m^2}{2} \Tr\left[U+ U^\dagger -2\right])}\right.\cr
&&\left.+ \frac{m^2 \Tr\left[U+ U^\dagger -2\right]}{2(1+\frac{\lambda m^2}{2} \Tr\left[U+ U^\dagger -2\right])}\right\}
+\frac{i}{4\pi}\int_{B^3} \frac{k}{4\pi} [\Tr U^{-1}d U]^3.\label{naaction}
\eea

{\em This is the 1+1 dimensional action we consider for the non-Abelian Heisenberg model}. In the next section 
we will see how to uplift it to 3+1 dimensions.
Upon expanding in small $\lambda$, and as before identifying $k=N_c$, we find now
\bea
S_M&\simeq& \int d^2x \left [\left( -\frac{N_c}{8\pi} X  +Z \right)+\frac{1}{16} \lambda \left(\left(\frac{N_c}{2\pi}\right)^2
(X^2- 2(\tilde X)^2)-16Z^2-8\frac{N_c}{2\pi}ZX\right)\right.\cr\
&&\left.+{\cal O}\left(\lambda^2\right)\right]
   + S_{WZ}\;,
\eea
where we denoted by $Z=\frac{m^2}{2}\Tr\left[U+ U^\dagger -2\right]$.
As was mentioned in the previous subsection, both $X$ and $X_{ij}^2$ are invariant 
under the full flavor chiral symmetry $U(N)_L\times U(N)_R$ transformations. 
However this is not a symmetry transformation of the term $Tr[ U + U^\dagger]$. This term is only invariant under  
the diagonal $U(N)$ transformation (with $A_L=A_R^\dagger=A$)
\be
U\rightarrow A U A^\dagger.
\ee

The corresponding current is 
\bea
{J^\mu}^A&=&\frac{i k}{4\pi}\left(\frac{\Tr\left[  U T^A \d_\mu U^\dagger +U^\dagger T^A\d_\mu U^\dagger\right]}{
\sqrt{1+ \frac{N_c}{2\pi}\lambda\left(1+\frac{\lambda m^2}{2} \Tr\left[U+ U^\dagger -2\right]\right) 
\Tr\left[\pa_\mu U(t,x)\pa^\mu U^\dagger(t,x)\right]}}\right.\cr
&&\left.+\epsilon_{\mu\nu}\Tr\left[  U T^A \d_\mu U^\dagger -U^\dagger T^A\d_\mu U^\dagger\right]\right).
\eea

In particular the diagonal $U(1)\subset U(N)$, which will play an important role below,
is  given by 
\bea
{J^\mu}&=&\frac{i k}{4\pi}\left(\frac{\Tr\left[  U  \d_\mu U^\dagger +U^\dagger \d_\mu U^\dagger\right]}{
\sqrt{1+ \frac{N_c}{2\pi}\lambda\left(1+\frac{\lambda m^2}{2} \Tr\left[U+ U^\dagger -2\right]\right) 
\Tr\left[\pa_\mu U(t,x)\pa^\mu U^\dagger(t,x)\right]}}\right.\cr
&&\left.+\epsilon_{\mu\nu}\Tr\left[  U \d_\mu U^\dagger -U^\dagger \d_\mu U^\dagger\right]\right).\label{Jbar}
\eea

Next we would search for soliton solutions of this deformed system.
The first point to notice is that for static configurations $X_{ij}^2$ vanishes, 
as does the WZ term, and  therefore the Lagrangian density for static  configurations takes the form 
\bea
{\cal L}_M &=& \frac{\left [ 1 - \sqrt{1+ \frac{k}{2\pi}\lambda\left(1+\frac{\lambda m^2}{2} \Tr\left[U+ U^\dagger -2\right]\right) 
\Tr\left[\pa_x U(x)\pa^x U^\dagger(x)\right]}\right ]  }{2\lambda(1+\frac{\lambda m^2}{2} \Tr\left[U+ U^\dagger -2\right])}\cr
&&+ \frac{m^2 \Tr\left[U+ U^\dagger -2\right]}{2(1+\frac{\lambda m^2}{2} \Tr\left[U+ U^\dagger -2\right])}.
\eea

We now take a particular ansatz for the group element $U(N)$
\be
U(x) =diag\left(1,1,...e^{-i\sqrt{\frac{k}{4\pi}}\phi(x)} \right)\;,
\ee

This ansatz, which takes us back from the non-Abelian setting to an Abelian one, will be relevant in the next subsection.
For this ansatz, we get that 
\bea
& & \frac{k}{4\pi} \Tr\left[\pa_\mu U(t,x)\pa^\mu U^\dagger(t,x)\right]= \frac12 \pa_\mu\phi\pa^\mu\phi  \CR
& & \half \Tr\left[U+ U^\dagger -2\right]= \cos\left(\beta\phi\right)-1 . \CR 
\eea

Substituting this into the action (\ref{naaction}),  we end up with the action of the 
$T\bar T$ deformation of the scalar field with a sine-Gordon potential.
In \cite{HNJS} the soliton solution for this action was derived and its properties 
were analyzed. We will describe some of them in the next subsection.

\subsection{Baryonic sector of bosonized $QCD_2$}

In \cite{Date:1986xe} it was shown that the low energy effective action of  multiflavor bosonized  
$QCD_2$, namely the theory with $SU(N_c)$ gauge symmetry and $U(N_f)$  flavor symmetry converges  in the strong limit to    
\be\label{QCD2}
S{sc}= N_c S_{WZW}(U) + m^2 \int d^2 x Tr[ U+ U^\dagger]\;, \qquad U\in U(N_f)\;,
\ee
where  the explicit expression for $m^2$ is given in \cite{Date:1986xe}. Thus, the action is 
a $U(N_f)$ WZW of level $k=N_c$, plus a potential term.

Solitons of this action should correspond to baryons of $QCD_2$. It was shown in 
\cite{Date:1986xe} that the lowest energy soliton is the one with the following structure of $U(x)$:
\be\label{Ux}
U(x) =diag\left (1,1,...e^{-i\sqrt{\frac{4\pi}{N_c}\phi(x)}} \right ) .
\ee

If we substitute this ansatz into (\ref{QCD2}) we get an action  for static configurations that takes the form
\be
S_s=\int d^2x \left[-\left (\frac{d\phi}{dx}\right )^2 -2 m^2\left( \cos\left ( \sqrt{\frac{4\pi}{N_c}}\right)-1\right ) \right ].
\ee

In the terminology of the sine-Gordon model we this action has $\beta=\sqrt{\frac{4\pi}{N_c}}$.

Alternatively, we could have followed the route taken in the previous subsection, of  
deforming the non-Abelian  action,   and substituting into it the form of $U(x)$ given in  (\ref{Ux}).

As was mentioned in the previous subsection, the $T\bar T$ deformation of the $T\bar T$ 
deformation of the  sine-Gordon action and its corresponding soliton  were analyzed in \cite{HNJS}. 

The soliton solutions for this action were derived in \cite{HNJS}. For our system, the soliton takes the form 
\be
\pm\sqrt{2}m \b(x-x_0)=\left.\left[\ln\left(\tan\left(\sqrt{\frac{\pi}{4N_c}}\phi\right)\right)+8\lambda m^2 
\cos\left(\sqrt{\frac{\pi}{N_c}}\phi\right)\right]\right|_{\phi(x_0)}^{\phi(x)}\;.
\ee

Using the expression for the baryon number current (\ref{Jbar}) we find 
that the  baryon number of this soliton solution is 
\be
Q_B=\int dx {J_B}_0 =\frac{N_c}{4\pi}\left(\phi(x\rightarrow \infty)- \phi(x\rightarrow -\infty)\right)=N_c \;,
\ee
as  it should be in the notation where each quark has unit baryon number.
 
Moreover in \cite{Date:1986xe} it was proven that the classical  mass of the soliton of 
the deformed theory is exactly the same as in the undeformed one which is given by\cite{Date:1986xe}
\be 
M_B= 4m\sqrt{\frac{4 N_c}{\pi}}= 4\left [N_c c m_q \left (\frac{g_c\sqrt{N_f}}
{\sqrt{2\pi}}\right ) \right ]^{\frac{1}{1+\Delta_C}} \sqrt{\frac{4 N_c}{\pi}}\;,
\ee
where $\Delta_c= \frac{N_c^2-1}{N_c(N_c+N_f)}$.


\subsection{ Boosting the 2d baryon to a  shock wave}

The strong coupling effective action of bosonized 2d QCD\cite{Frishman:2010zz} does not admit a 
shock wave solution. This follows from the fact  that the necessary condition to have such a 
solution\cite{Nastase:2015ixa}, namely to have an infinite tower of higher derivative terms in the action,  is not fulfilled in that case.
However, as was shown in \cite{Conti:2018jho} and more explicitly in \cite{HNJS} 
the deformed sine-Gordon action does admit a  shockwave behavior. Hence, following the 
the discussion above also the deformed $QCD_2$ with the particular ansatz of the group 
element, also  admits a shockwave behavior. Thus, the deformed  $QCD_2$ action is a framework 
in which one can discuss the static properties of the baryonic soliton but also its 
scattering in the form of shockwave collisions.


\section{Boosted Skyrmion-like solitons as shockwaves}

As we saw in the section 3, the shockwave solution can be used to describe the (saturation limit for the cross section for the)
collision of high-energy nucleons. The nucleons, together with the pion field they generate, when 
boosted will become first pancake-like, then delta 
function sources surrounded by shockwave pion field, for which the Heisenberg shockwave is a model. 

On the other hand, there is one way to deal with nucleons {\em at rest}
not as sources for the pion field, but rather as sourceless topological pion field configurations, 
representing both the nucleon and the surrounding pion field, as solutions for the low energy expansion in QCD (chiral perturbation theory). 
That is the Skyrme solution for the nonlinear sigma model (for the $SU(2)$-valued field corresponding to the 
three physical pions, $\pi^\pm$ and $\pi^0$)
with a Skyrme term. It is also known that a topological solution still exists when we replace the Skyrme term with other higher order terms in the action. 

It is a reasonable hope then that by boosting a Skyrme-like solution we should get a shockwave of the 
type found by Heisenberg (at least in the region when there is a single wave), as it was argued also in 
\cite{Kang:2004jd,Kang:2005bj,Nastase:2005pb,Nastase:2008hw}. Here we will show that by proposing an action that is a natural nonabelian 
generalization of the $T\bar T$ action we obtain a Skyrme-like solution that when boosted looks like Heisenberg's shockwave, near $s=0$ and for $
t<0$.

\subsection{Abelian analogue of solitons and boosting}

In this subsection, after reviewing previous results on various solitons, we will argue for the existence of a soliton of the $T\bar T$ 
action that can pass smoothly through the $1-\lambda V=0$ singularity, and then argue about what happens when we infinitely 
boost such a solution.

\subsubsection{BIons and catenoids}

We start this subsection by reviewing the previous attempt for the Abelian scalar theory, in \cite{Nastase:2005pb}, 
to realize the above program. 
As we said, the hope was initially that a single scalar action would generate both the shockwave-like solution 
and the Skyrmion-like solution, so that we can we could infinitely boost the latter into the former. 

The first observation is that, by restricting to an abelian analog (a single real scalar), the Skyrme-like solution 
would be a solution of finite energy, since 
the nucleon represented as a Skyrmion has a finite energy also. There is one such solution, the BIon solution 
found by Born an Infeld in 1934 in 
\cite{Born:1934gh}, in their attempt to replace the electron solution of classical electrodynamics, which has 
an infinite field energy (the infinite energy 
shows the need to go to quantum field theory, i.e., to QED, but Born and Infeld sought a {\em classical} replacement). 

The Born-Infeld Lagrangian (putting the relevant length scale $l$ to 1), 
\bea
{\cal L}&=&-\sqrt{-\det\left(\eta_{\mu\nu}+\frac{F_{\mu\nu}}{\sqrt{2}}\right)}
=-\sqrt{1+\frac{F_{\mu\nu}F^{\mu\nu}}{2}-\left(\frac{F_{\mu\nu}*F^{\mu\nu}}{2}\right)^2}
\cr
&=&-\sqrt{1-\vec{E}^2+\vec{B}^2-(\vec{E}\cdot \vec{B})^2}\;,
\eea
where in the last form we wrote the electric and magnetic fields, has the 
equation of motion in the presence of a static source with charge density $\rho$
(so zero magnetic field) of 
\be
\vec{\nabla}\cdot \vec{D}=\rho\;,
\ee
where as usual we defined 
\be
\vec{D}=\frac{\d {\cal L}}{\d \vec{E}}.
\ee

The electric field $\vec{E}$ is then finite ($\leq 1$), since we obtain 
\be
\vec{E}=-\frac{\vec{D}}{\sqrt{1+\vec{D}^2}}.
\ee
In terms of $\vec{E}=-\vec{\nabla}\phi$, with $\phi$ the electric potential, the Lagrangian on the time-independent solution is 
\be
{\cal L}=-\sqrt{1-(\vec{\nabla}\phi)^2}\;,
\ee
and the solution for point-like source $\rho=q\delta^3(r)$ is
\be
\phi(r)=q\int_r^\infty \frac{1}{\sqrt{q^2+x^2}}\;,
\ee
and has {\em diverging energy density at $r=0$ (where $|\vec{E}|=|\phi'(r)|=1$), but finite energy}, 
\be
E_\phi=4\pi \int_0^\infty r^2 dr \left[\frac{1}{\sqrt{1-\phi'(r)^2}}-1\right]=4\pi q^{3/2}\int_0^\infty\frac{dx}{x^2+\sqrt{x^4+1}}=4\pi q^{3/2}\frac{\Gamma[1/4]^2}{6
\sqrt{\pi}}.
\ee

Note that the on-shell Lagrangian looks like one for a scalar DBI action, but {\em with the wrong sign inside the square root} (and there is no 
corresponding time derivative term, since the scalar is not a real scalar, but is part of a vector). 

On the other hand, the shockwave-like solution can be understood as a solution with a jump in the derivative for a true scalar theory, as Heisenberg did, 
which (as Heisenberg also did) restricts us to the DBI action, 
\be
{\cal L}=-\sqrt{-\det\left(\eta_{\mu\nu}+\d_\mu X \d_\nu X\right)}=-\sqrt{1+(\d_\mu X)^2}. 
\ee

If we want a static solution for it, define (similarly to the Born and Infeld case) $\vec{F}=\vec{\nabla}X$, so the on-shell Lagrangian on this ansatz is 
\be
{\cal L}=-\sqrt{1+\vec{F}^2}.
\ee

The equation of motion in the static case with a source $\rho$ is 
\be
\vec{\nabla}\cdot \vec{C}=\rho\;,
\ee
where we have defined 
\be
\vec{C}=\frac{\d {\cal L}}{\vec{F}}.
\ee

Then 
\be
\vec{F}=\frac{\vec{C}}{\sqrt{1-\vec{C}^2}}.
\ee

The solution for point-like source $\rho=q\delta^3(r)$ is the "catenoid", 
\be
X(r)=q\int_r^\infty\frac{dx}{\sqrt{x^4-q^2}}\;,
\ee
which however, as we see, has a "horizon" at $r_0=\sqrt{q}$, where $X(r)$ is finite but $X'(r)$ diverges. When we think of the DBI action as the 
action for the position of a D-brane, this is understood as one half of a solution of D-brane-anti-D-brane, connected by a funnel (solution obtained by 
adding a mirror image of the catenoid). The solution also has a diverging energy density, yet a finite energy, but now at this "horizon", 
\be
E_X=4\pi \int_{r_0}^\infty r^2 dr\left[\sqrt{1+X'^2}-1\right]=4\pi q^{3/2}\int_1^\infty x^2 dx\left[\frac{x^2}{\sqrt{x^4-1}}-1\right].
\ee

Note that both BIon and catenoid solutions become at large $r$ just 
\be
\phi (r)\simeq \frac{q}{r}.
\ee

Perhaps one can find a boosting limit, where one takes $q\rightarrow 0$, so as to have no observable horizon with diverging energy density, 
but keeping $q\gamma$ finite in the limit (where $\gamma$ is the relativistic factor $1/\sqrt{1-v^2}$), just like when we infinitely boost the Schwarzschild 
metric we obtain the Aichelburg-Sexl shockwave metric \cite{Aichelburg:1970dh}, 
by keeping $p=M\gamma$ finite in the limit where $M\rightarrow 0, \gamma\rightarrow \infty$), 
but we have not been able to show this. In any case, we will see that in the nonabelian case, the generalization with this positive sign inside the 
square root doesn't have a Skyrmion-like solution, whereas the one with the negative sign does. 

Next, it was found that in the D-brane action, both $X$ and $\phi$ are present, as
\be
S_{\rm DBI}=-\int d^4x \sqrt{[1-(\vec{\nabla} \phi)^2][1+(\vec{\nabla}X)^2]+\vec{\nabla}\phi\cdot
\vec{\nabla}X}.
\ee
Therefore this action has both BIon (in $\phi$) and catenoid (in $X$) static solutions. One can hope then that perhaps by infinitely boosting a more general, 
BIon plus catenoid solution, one can obtain the shockwave solution, but again that is not clear (and we still have the problem of the nonabelian 
generalization). 

\subsubsection{BIon-like solution from $T\bar T$ action}

This was the case so far. But with the $T\bar T$ action replacing the DBI action (with mass term inside the square root, which is irrelevant 
for the solution near $s=0$), one more possibility arises. For the $T\bar T$ action, we have a single scalar $\phi$.
We can have a BIon-like solution, namely, a solution of finite energy defined for all $r$ until 0, and yet be defined in terms of the same scalar field $X$
(now called $\phi$, but meaning a true scalar, not the electric potential). And then infinitely boosting it we can obtain the shockwave solution. 

The catenoid solution has $\phi'(r)\rightarrow \infty $ at some $r=r_0$, but $\phi(r_0)$ finite, but to get to $r=0$ with a finite energy like for the BIon, we need 
the opposite sign inside the square root. However, that can happen at large field $\phi$, since the coupling of the $(\d_\mu\phi)^2$ term is 
$\bar \lambda =\lambda (1-\lambda V)$. Then we expect to have both catenoid-like and BIon-like solutions. Note that the positive sign of $\bar \lambda$ 
is associated not just to the catenoid solution, but also to the shockwave solution. 

Both catenoid-like and Bion-like go to $q/r$ at $r\rightarrow \infty$, since then also $V\simeq 0$, so $\bar \lambda\simeq \lambda$. 
As we go to smaller $r$, they start to differ. Then $\phi$ becomes large, and then so 
$V(\phi)$, at least in the case that we have only a mass term (and maybe also a $\phi^4$ term), so the question is whether $\bar \lambda$ reaches zero 
or not. If it stays positive, we have a catenoid-like solution (and also shockwave solution, in the boosted case), which develops a horizon at $r_0\simeq 
\sqrt{q_0}$, where $\phi'(r_0)=\infty$, $\phi(r_0)$ finite. You might think, but the catenoid is 
singular, so how come for him, $V$ can stay small enough? The answer is that the singularity is in 
the {\em derivative}, not the field; the field can stay small. 
If $\bar\lambda$ reaches zero, and changes sign, we have a BIon-like solution (which stands for the Skyrmion-like one in the nonabelian case). 

The question we need to  answer   is whether it is possible to have a 
continuous solution, for the BIon-like case, that (must!) change sign for $\bar \lambda$ from $\bar \lambda>0$
at $r\rightarrow \infty$ ($\phi$ small) to $\bar \lambda<0$ at $r\rightarrow 0$ ($\phi$ large). 
If we can, then we should be able to infinitely boost such a BIon-like solution to the shockwave solution, something that seemed impossible before. 

To see whether we can have a continuous solution for the BIon-like case, we check whether the ansatz for having a {\em finite} (of order 1, neither zero 
nor infinity) $\phi'(\phi_0)$ for $\phi_0$ a {\em finite} value for which $\bar \lambda=0$ is consistent for  the solutions of  the equations of motion. 

Consider the value $\phi=\phi_0$ of the field for which $\lambda V(\phi_0)=1$ (so that $\bar \lambda 
=0$). Then 
\be
1-\lambda V(\phi)\simeq -\lambda V'(\phi_0)(\phi-\phi_0)=-\lambda V'(\phi_0)\delta \phi\;,
\ee
where we have defined $\delta\phi\equiv\phi-\phi_0$. Then we find that the Lagrangian on the above ansatz becomes
\be
-{\cal L}\simeq\frac{1+\sqrt{1-8 \lambda^2V'(\phi_0)\delta \phi (\delta \phi')^2}}{-2\lambda^2V'(\phi_0)
\delta\phi}.\label{Lapprox}
\ee

Its equation of motion, $2\lambda^2V'(\phi_0)\delta \phi\times \delta S/\delta (\delta \phi)$, 
is
\bea
&&-\frac{1}{\delta \phi}\left(1+\sqrt{1-8 \lambda^2V'(\phi_0)\delta \phi (\delta \phi')^2}\right)
-\frac{4\lambda^2V'(\phi_0)\delta\phi'^2}{\sqrt{1-8 \lambda^2V'(\phi_0)\delta \phi (\delta \phi')^2}}\cr
&&+2\lambda^2V'(\phi_0)\delta \phi\left[\frac{4\delta \phi'}
{\sqrt{1-8 \lambda^2V'(\phi_0)\delta \phi (\delta \phi')^2}}\right]'=0.
\eea

Multiplying with $\delta \phi \sqrt{1-8 \lambda^2V'(\phi_0)\delta \phi (\delta \phi')^2}$, and 
after some (longish) algebra, we get
\bea
&&-1+12 \lambda^2 V'(\phi_0)\delta\phi\delta\phi'^2+8\lambda^2V'(\phi_0)\delta \phi^2\delta \phi''\cr
&&+[1-8\lambda^2V'(\phi_0)\delta \phi \delta\phi'^2]^{3/2}=0.
\eea

The ideal solution is when there is nothing special at this point $\phi_0$, reached at position $x=x_0$, 
meaning that we can have a regular Taylor expansion of the field $\phi$, 
\be
\delta \phi(x)=A\delta x=\phi'(x_0)(x-x_0)\;,
\ee
with $A=\phi'(x_0)$ a nonzero and finite constant. Substituting this ansatz into the above equation, the leading 
term near $x=x_0$, is $-1+1=0$, therefore the equation is consistent! That means that the ansatz above was correct (if it was 
incorrect, we could have tried instead, for instance, $\delta \phi(x)=A \delta x^\a$, with $\a$ a power less than 
1, and maybe then we obtained consistency, 0=0, for some $\a$; as it is, we had the correct ansatz).

Finally, that means that we {\em can} find a BIon-like solution, for which the derivative of the field at the point that $\bar\lambda$ changes sign 
stays finite. 

Note that we have really considered the approximate Lagrangian (\ref{Lapprox}) in order to obtain the solution. 
But this Lagrangian is not necessarily obtained from the $T\bar T$ deformation Lagrangian, but can come from any Lagrangian of the 
type
\be
-{\cal L}=\frac{h(\phi)+\sqrt{1+2\lambda(1-\lambda \tilde g(\phi))(\d_\mu\phi)^2}}{2\lambda(1-\lambda \tilde g(\phi))}\;,
\ee
if for $1-\lambda \tilde g(\phi_0)=0$, then $h(\phi)=1$. The notion of having a {\bf potential} $V(\phi)$, and of $\tilde g(\phi)$ 
to be related to $h(\phi)$ is not needed. 

In this subsection, we considered a single scalar, but by the embedding of the single scalar action into the nonabelian one, 
we know that the same conclusions can be reached about the latter. The BIon-like solution becomes a Skyrmion-like solution, 
and the shockwave still a shockwave. Thus, we have shown that there is a  potential for the Skyrmion-like solution to 
be boosted to a shockwave. Exactly how that is done will be examined next, but the existence of a horizon hiding a 
region with wrong sign inside the square root (which allows for a topological solution, of finite energy), 
yet which can be scaled down to zero in a certain limit, is essential.

\subsubsection{Infinitely boosting and indirect limit}

Next, we need to understand what is the condition needed to show that the infinitely 
boosted BIon-like solution (which we still don't know, we just proved it exists)
goes to the shockwave solution. To do that, we consider how we know that an infinitely 
boosted Schwarzschild black hole is the Aichelburg-Sexl schockwave. 

In  the paper of Aichelburg and Sexl \cite{Aichelburg:1970dh}, they took a nontrivial limit on the Schwarzschild solution 
for a black hole, which has a singularity (for the point mass $M$) at $r=0$ and a horizon at $r_h=2M$. The horizon exists because
the solution is written in terms of the harmonic function (in 3 spatial dimensions)
$f(r)=1-2M/r$. The Einstein equations reduce to the Poisson equation for $f(r)$. 
In the nontrivial limit of $M\rightarrow 0$, $\gamma=1/\sqrt{1-v^2}\rightarrow \infty$, 
but $p\simeq M\gamma=$finite (for $v\rightarrow 1$), we obtain the Aichelburg-Sexl shockwave. 

A general pp wave would be 
\be
ds^2=2dx^+dx^-+dx_idx_i +(dx^+)^2H(dx^+,x_i).
\ee

For a shockwave, we have 
\be
H(x^+,x^i)=\delta(x^+)\Phi(x_i).
\ee

For the Aichelburg-Sexl shockwave, we have $\Phi(x_i)$ harmonic in 
the transverse dimensions, with Poisson source $p$, so in $D=4$ dimensions, 
\be
\Phi(x_i)=-4G_{N,4}p \ln r^2\;,
\ee
where $r^2=x_ix_i$, and in general
\be
\Phi(x_i)=\frac{16\pi G_{N,d}}{\Omega_{d-3}(d-4)}\frac{p}{r^{d-4}}.
\ee 

In previous works, in particular in \cite{Kang:2005bj}, it was argued that for a similar A-S shockwave in the background of the 
gravity dual to QCD (say, in cut-off $AdS_5$), $\Phi(x_i)$ should be associated with the pion field profile, at least at large $r$, which means that 
what is true for obtaining the A-S shockwave, should also be true for obtaining the pion shockwave, which is why we are reviewing the gravitational 
shockwave limiting procedure. 

The limiting boost procedure in the A-S paper is complicated, and also involves a certain nontrivial change 
of coordinates, but one can easily get the qualitative 
picture without calculations: since $M\rightarrow 0$, the horizon goes to zero, $r_h\rightarrow 0$, which is 
why there is no horizon in the shockwave. Then anything 
gravitational (curving space) moving at the speed of light must be a pp wave, 
and if it is point-like (like the mass in the Schwarzschild metric is), it is a 
shockwave (delta function in $x^+$). 

But instead of the complicated limiting procedure, one can use a shortcut, and find the same 
result in an easier way. Instead of taking the limit on the metric, 
we take it on the equation it satisfies. We can check (and find also without explicit calculations) 
that the only nonzero component of the Ricci tensor $R_{\mu\nu}$ 
for the pp wave metric is $R_{++}$. Then this must be proportional to $g_{++}$, and 
(since $R_{\mu\nu}$ has 2 derivatives), we must have $R_{++}\propto
\d_i\d_i g_{++}$ (because of general coordinate invariance, it cannot have $\d+\d_i$, or $d_+^2$, 
the only other possibilities for 2 derivatives acting on $g_{++}=
H(x^+,x_i)$). In fact, with the correct coefficient,
\be
R_{++}=-\frac{1}{2}\d_i^2H(x^+,x_i)\;,
\ee
which in fact equals the linearized result (it is the only case we know of, when the Einstein equation is exactly equal to its linearization). 
The Einstein equation makes this equal to $T_{++}$. 

Now a point particle moving at the speed of light has only $T_{++}$, 
and moreover has momentum $p$, as we defined, so 
\be
T_{++}=p\delta^{d-2}(x_i)\delta(x^+)\;,
\ee
leading in turn to the shockwave ansatz $H=d\delta(x^+)\Phi$, 
which finally leads to 
\be
\d_i^2\Phi(x_i)=-16\pi G_{N,d}p \delta^{d-2}(x_i)\;,
\ee
with solution the A-S shockwave. 
Thus after the infinite boost, the Einstein equations reduced to the equation for the field $\Phi(x_i)$ in the pp shockwave ansatz. 
We see that the pp wave ansatz is a very interesting one: the usually very nonlinear Einstein equations linearize 
on the ansatz (the linearized equation is actually exact). 

We want to mimic the procedure in the case of the abelian pion action. In this case, there is no reduction of the degrees of freedom by the ansatz 
($g_{\mu\nu}$ in the general Einstein equations, $\Phi$ on the ansatz). Instead, the 
variable is a single real scalar $\phi$ before and after the infinite boost. 

Then all we need to show is that the equations of motion, or equivalently, the Lagrangian, for the BIon-like ansatz, go over to the Lagrangian for the 
shockwave, under the inifinite boost. 

First, we need to understand the limiting procedure. To do that, we go back to the shockwave energy-momentum tensor (\ref{enmomten}), to find 
the total momentum of 
one of the colliding shockwaves, for instance $x^-=0$. For the A-S gravitational shockwave on $x^+=0$, only $T_{++}$ was nonzero, 
but it was a constant times a delta function, integrating to 1, so finite total momentum. 
Yet now, as we said, $x^-=0$ implies only $T_{--}$, which is a large, yet finite constant, times $ (d\phi/ds)^2 (x^+)^2 $. Since also $(x^+)^2$ 
is a constant, we have (since $\phi \sim \sqrt{s}$)
\be
T_{--}\sim \left(\frac{d\phi}{ds}\right)^2\sim \frac{1}{s}\propto \frac{1}{\sqrt{x^+x^-}}.
\ee
By integrating over $x^-$, this means that there is a vanishing momentum located at $x^-=0$. 

On the other hand, using 
\be
\dot \phi \equiv \frac{\d \phi}{\d t}=\frac{d\phi}{ds}\frac{t}{s}\;,
\ee
the energy is (ignoring the finite and vanishing terms in the energy-momentum tensor)
\bea
E&=&\int dx T_{00}\simeq \int dx \frac{\dot \phi^2}{\sqrt{1+l^4(\d_\mu\phi\d_\nu\phi g^{\mu\nu}+m^2\phi^2)}}\cr
&\propto& \int dx
\left(\frac{d\phi}{ds}\right)^2\frac{t^2}{s^2}\propto \int dx \frac{1}{s^3}=\int dx\frac{1}{(t-x)^{3/2}(t+x)^{3/2}}\;,
\eea
and integrating over $x$ at fixed $t$, around $x=t$, we get a divergence. So the total energy of the field is divergent, which 
is consistent with a object of finite mass $M$ (the Skyrmion) boosted with an infinite $\gamma$. But that means that now 
$M\gamma$ is not kept finite, but is still infinite. 

In fact, now we realize $M$ {\em should} be kept fixed in the limit. Indeed, $M$ can only depend on the parameters of the theory, and 
those are fixed by experiment, and cannot be varied. 

Now the limit only makes use of Lorentz contraction: the horizon in the $x$ direction has zero size now. 
But the action on the BIon-like ansatz included $\phi'(r)$ and $\phi(r)$, where $r$ was the total radial coordinate, which is now 
(since $v\simeq 1$)
\be
r^2\simeq [(x-t)\gamma]^2+\tilde r^2=(x^-)^2\gamma^2+\tilde r^2\;,
\ee
where $\tilde r$ represents now the transverse coordinate. That means that the dependence on $\tilde r$ becomes negligible. 
Of course, now also $\dot \phi$ becomes nonzero in the ansatz. Thus trivially, we have that the equations of motion on the BIon-like ansatz 
become the ones on the shockwave ansatz under the infinite boost, therefore the 
BIon-like solution {\em should } become the shockwave one
(in the single wave region, $t<0, x<0$, for instance), as wanted. 

It is just that we can only "see" the outside horizon solution (with $\bar \lambda >0$), the "inside horizon" part has become obscured.  
We can ask: where is this "inside horizon" region? Its extent in the $x$ direction has become 
zero by the infinite boost, that is why we don't see it. 
Of course, its extent in the transverse directions has remained the same, but it is situated 
{\em on the singularity} $s=0$, so it is impossible to reach.

\subsection{A nonabelian version of $T\bar T$-like actions and chiral perturbation theory}

In this subsection, after reviewing Skyrmion-like solitons of non-Abelian actions and chiral perturbation theory, we will
propose a 3+1 dimensional, non-Abelian version of the $T\bar T$ action, 
and argue that it consistent with chiral perturbation theory confronted by experiment.

\subsubsection{Review of Skyrmion-like solutions in non-Abelian generalizations, and chiral perturbation theory}

In \cite{Nastase:2005pb}, one of us has considered the possibility of a certain nonabelian generalization of the D-brane action that can have 
Skyrmion-like solutions. The reason why we considered the D-brane action is that, as we saw, the Skyrmion-like solution is the analog of the 
BIon-like solution. 

Indeed, in a paper by Pavlovsky \cite{Pavlovsky:2002bx}, 
it was shown that we can have Skyrme-like solutions for a nonabelian DBI-like scalar 
action, {\em but with the wrong sign inside the square root}. 

For effective actions for QCD in the low energy expansion
(note that the various nonlinear sigma models used for QCD are, at least for the relevant first order
in chiral perturbation theory, all equivalent through field redefinitions, etc.), one way of defining the 
theory is by an $SU(2)$ matrix $U$ containing the linearized pions $\pi^i \sigma_i$
in an expansion of $U$ around the identity, from which we form the quantities
\be
L_\mu=U^{-1}\d_\mu U.
\ee

Then the  action  of \cite{Pavlovsky:2002bx} is 
\be 
{\cal L}_P=\Tr\sqrt{1+\frac{1}{2\b^2}L_\mu L^\mu}.
\ee

Considering the spherically symmetric ("hedgehog") ansatz
\be
U=e^{iF(r)\vec{n}\cdot \vec{\sigma}}\;,
\ee
where $\vec{n}=\vec{r}/r$ is the radial unit vector and $\sigma_i$ are the Pauli matrices), we find 
\bea
L_i&=& in_i \vec{n}\cdot \vec{\sigma}\left(F'-\frac{\sin F \cos F}{r}\right)+\frac{i\sigma_i}{r}\sin F \cos F-i\frac{\sin^2F }{r}\epsilon_{ijk}n_j 
\sigma_k\Rightarrow\cr
L_i^2&=& -F'^2-2\frac{\sin^2F}{r^2}-4\vec{n}\cdot \vec{\sigma}\frac{\sin^3F \cos F}{r^2}\Rightarrow\cr
\Tr[L_i^2]&=&2\left[-F'^2-2\frac{\sin^2F}{r^2}\right].
\eea

Now, both in  \cite{Pavlovsky:2002bx} (implicitly, from the equations deduced from it) and 
in  \cite{Nastase:2005pb} (explicitly), the term with $\vec{n}\cdot \vec{\sigma}$ was erroneously forgotten, so that 
it was claimed that on the hedgehog ansatz, the $L_\mu L^\mu$ appearing inside the square root was already proportional 
to the identity. For all the calculations in these papers to be valid, however, we need then to change the Lagrangian to
one with square root inside,  
\be
{\cal L}'_P=\sqrt{1+\frac{1}{4\b^2}\Tr[L_\mu L^\mu]}\;,
\ee
since then the Lagrangian on the hedgehog ansatz is unmodified with respect to the results there, and is
\be
{\cal L}'_P=\sqrt{1-\frac{1}{2\b^2}F'^2-\frac{\sin^2 F}{\b^2 r^2}}.
\ee

We note then that indeed, the sign inside the square root is wrong (minus, instead of plus) with respect
to the scalar DBI action. But, together with the wrong sign in front of the action, this means that 
at low fields, the action still becomes the same canonical scalar, and only the higher nonlinear terms 
have different signs. 

This  action\cite{Pavlovsky:2002bx} has a Skyrmion-like solution, which can be also thought, comparing with the 
D-brane action above, as a BIon-like solution (since $F(r)$ is like $\phi(r)$, it has the same sign 
inside the square root). But, since the sign inside the square root is wrong, we can say (and prove
rigorously, see  \cite{Nastase:2005pb}) that it {\em doesn't} have a catenoid-like solution (with a 
``horizon''), and so also not a shockwave solution. The Skyrmion solution has $F(0)=N\pi$ and $F(\infty)=0$, because of the 
topological constraint, which means that the energy and baryon number of the solution is quantized.

The baryon number density on the Skyrmion solution can be defined. On any solution defined in terms of 
$U= e^{iF(r)\vec{n}\cdot \vec{\sigma}}$, we have the baryon number (winding number, or topological number)
\be
B=-\frac{1}{24\pi^2}\int d^3 r \epsilon^{ijk}\Tr[U^{-1}\d_i U U^{-1}\d_j U U^{-1} \d_k U]\;,
\ee
so the baryon number density is what is inside $\int d^3 r$. Since we get 
\be
B=-\frac{1}{2\pi^2} \int d^3 r \; F'(r)\frac{\sin^2 F(r)}{r^2}=-\frac{2}{\pi}\int_0^\infty dr \; F'(r)\sin^2 F(r)\;,
\ee
it means that the baryon number density is 
\be
\rho_B(r) =-\frac{1}{2\pi^2}F'(r)\frac{\sin^2 F(r)}{r^2}.
\ee

Then, if $F(0)=\pi)$, we have a baryon, and if $F(0)=-\pi$, we have an anti-baryon. Moreover, we can see in the density, that for the 
baryon, $F'(r)<0$ ($F$ decreases until $F=0$ at infinity), so $\rho_B(0)>0$, whereas for the antibaryon $F'(r)>0$ ($F$ increases until 
$F=0$ at infinity), so $\rho_B(0)<0$.

But boosting nucleons should lead, according to Heisenberg, to pion 
shockwave solutions. But nucleons are described by Skyrmions, and so perhaps Skyrmions should 
be boosted to shockwaves. But in the case described in \cite{Pavlovsky:2002bx}, we have one, but not the other solution, 
while in the DBI D-brane action, we have both solutions, but in terms of different {\em fields}.

As we saw, the DBI D-brane action is Abelian, whereas the Skyrme-type Pavlovsky one is non-Abelian, 
which is essential in order to obtain a topological Skyrme-type solution, so we need to write a 
non-Abelian generalization. In \cite{Nastase:2005pb} an attempt at generalization was made only for 
the space dependence (without the 
time dependence), and only for $|\vec{\nabla}\phi|\ll \bar M$ (energies below some cut-off), so 
the fields are $A_0=\phi$, or more precisely $\vec{E}=\vec{\nabla}\phi$, and 
$L_i=U^{-1}\d_i U$ replacing $X$.
The action was 
\be
\frac{\cal L}{\rm const.}=1-\Tr\sqrt{\left(1-\frac{\vec{E}^2}{M_2^2}\right)\left(1-\frac{L_i^2}{M_1^2}
\right)-\frac{(E_iL_i)^2}{M_1^2M_2^2}+\frac{L_0^2}{M_1^2}+\frac{m_\pi^2}{M_1^2}(U+U^\dagger
-2)+M_A^2\phi^2}\;,
\ee
but it is not very satisfying. It was found that it has a BIon-like solution {\em in terms of $\phi$}, 
and at $\phi=0$ it has a catenoid-like solution (with a ``horizon'') in terms of $F$, but no Skyrmion-like 
solution for $F(r)$, since it has the wrong sign with respect to \cite{Pavlovsky:2002bx} (the correct sign is for $\phi$, 
but that cannot have the Skyrmion radial ansatz, since it is not in the fundamental representation of 
$SU(2)$).

But as we said, one should rather have an action that corresponds to a particular form of chiral perturbation theory. 
To see what that means, we review a few relevant points about the latter.


The Lagrangian for low-energy QCD contains the $u$ and $d$ quarks (and maybe the $s$ quark), and some mass terms for the 
quarks ($u$ and $d$ certainly, $s$ if we include it), and in the massless quark limit  is invariant under global $U(2)_L\times U(2)_R$
(extended to $U(3)_L\times U(3)_R$ if $s$ is included). The masses break the symmetry, so the Goldstone bosons for the 
symmetry breaking, the pions $\pi^a=(\pi^+,\pi^-,\pi^0)$ become pseudo-Goldstone bosons. 

The diagonal part of the $U(1)_L\times U(1)_R$ component is 
baryon number, as used in section 4, and the other $U(1)$ is a chiral symmetry broken by anomalies. The $SU(2)_L\times SU(2)_R$ 
is broken spontaneously to the diagonal $SU(2)_V$, and the other $SU(2)$ is a chiral symmetry, that must be broken spontaneously, 
but its mechanism in QCD is unknown. 

To deal with this spontaneous breaking phenomenologically, one constructs a low energy 
Lagrangian in terms of the low energy physical (gauge invariant) mesonic states $\pi^a$ 
and $\sigma$, put together into the matrix ($\tau^a$ are the Pauli matrices)
\be
\Sigma=\sigma\one +i\tau^a \pi^a\;,
\ee
and constructs the {\em linear sigma model} Lagrangian for it, similar to the Higgs Lagrangian. This restricts $\Sigma$ to its 
vacuum value $v$, if we ignore the Higgs-like fluctuation $s(x)$, 
\be
\Sigma=(v+s(x))U(x)\;,\;\;\;
U(x)=e^{i\tau^a\pi'^a(x)/v}.
\ee

This results in the {\em NonLinear Sigma Model } for the unitary matrix $U$, $U^\dagger U=\one$, 
\be
{\cal L}_{NL}=-\frac{v^2}{4}\Tr[\d_\mu U \d^\mu U^\dagger].\label{NL}
\ee

This is however only the lowest term (second order) in a low energy expansion in derivatives $\d_\mu$, or momenta $p_\mu$, 
which is usually taken together with pion masses $m_\pi$, because of the on-shell relation $p^2=m_\pi^2$. 
The next term would be a $p^4$ term, so 
\be
{\cal L}={\cal L}_2(p^2)+{\cal L}_4(p^4)+{\cal L}_6(p^6)+...
\ee

The NLSM Lagrangian (\ref{NL}) is the most general chiral invariant effective Lagrangian with 2 derivatives written in terms of the 
unitary $U$, however, there are many way to re-express it via field redefinitions, for instance as an $SO(4)$ vector model, etc. 

Since in this work we have dealt only with unitary matrices $U$, we consider it to be the basic first term in the expansion. Indeed, 
it is the first term in the expansion in derivatives of the non-Abelian version of the Heisenberg or $T\bar T$ deformation, 
see for instance (\ref{naaction}). 

At the next order ($p^4$) however, there are many possible terms. If we also consider the coupling to external fields (that 
makes the global symmetries local, by considering a covariant derivative), adding the terms
\be
\bar q\gamma^\mu\left(V_\mu +\gamma_5 A_\mu\right)q -\bar q(s-i\gamma_5p)q
\ee
to the QCD low energy action, we find corresponding terms that can be added to the effective action, now written in terms of 
$U$, $\chi=s+ip$, and also in terms of $a^L_\mu$ and $a^R_\mu$, equal to $V_\mu\pm A_\mu$, with their corresponding 
field strengths, $f_{\mu\nu}^L$ and $f_{\mu\nu}^R$.

\subsubsection{Nonabelian action with Skyrmion-like solutions and comparison with chiral perturbation theory}

We now will write a nonabelian version of the $T\bar T$ action extended to 3+1 dimensions, 
from the point of view of chiral perturbation theory, 
and then we will compare with the general expectations about the latter discussed in the 
review \cite{Scherer:2002tk}. That means that we want to 
write an action in terms of the matrix $U$ above, and the associated $L_\mu=U^{-1}\d_\mu U$. 

To do that, we note the standard replacement, when going from one scalar to the $SU(2)$-valued scalars $U$, 
\be
(\d_\mu\phi)^2\rightarrow -L_\mu L^\mu\;,
\ee
that leads one from the free massless scalar to the nonlinear sigma model action 
\be
{\cal L}_{\rm NLSM}=\frac{f_\pi^2}{4}\Tr[L_\mu L^\mu]=-\frac{f_\pi^2}{4}
\Tr[\d_\mu U\d^\mu U^{-1}].
\ee

But this is the starting point of the $T\bar T$ deformed action in 1+1 dimensions considered in section 4, trivially extended 
to 3+1 dimensions. 
Moreover, as we argued there, we need to also add a WZW term, in order for the Noether charge to match the topological 
charge, and be equated to the baryon charge. That means that we extend to 3+1 dimensions the analysis of section 4. 

Then at zero potential ($V=0$), the Abelian $T\bar T $ action (now equal to DBI) extended to 3+1 dimensions and generalized to a
nonabelian action relevant for QCD in the low energy expansion is 
\be 
{\cal L}_{DBI,NLSM}=-f_\pi^2 \b^2\left[\sqrt{1-\frac{1}{4\b^2}\frac{N_c}{2\pi}\left(\Tr[L_\mu L^\mu]-\frac{1}{4\b^2}\frac{N_c}{4\pi}
\tilde X^2\right)}-1\right]+{\cal L}_{WZW}\;,\label{DBINL}
\ee
which has the opposite sign inside the square root with respect to the Pavlovsky case (that 
has a Skyrmion solution, as we saw), but otherwise it is 
the same. We need to define $\tilde X^2$ and ${\cal L}_{WZW}$ in 3+1 dimensions. The latter is standard, and the former 
can be extended most naturally (so as to go back to the 1+1 dimensional result upon dimensional reduction) as 
\be
\tilde X_1^2=\frac{1}{2}\epsilon_{\mu\nu\rho\sigma}\epsilon_{\lambda \tau\rho\sigma}\Tr[L_\mu L_\lambda]\Tr[L_\nu L_\tau]\;,
\ee
which then gives a nonzero contribution {\em on the static ansatz} $L_0=0$, 
\bea
\tilde X_1^2&=&\epsilon_{imp}\epsilon_{jnp}\Tr[L_i L_j]\Tr[L_m L_n]\cr
&=&8\frac{\sin^2 F}{r^2}\left\{2F'^2+\frac{\sin^2F}{r^2}+\frac{\sin^4 F \cos^2 F}{r^2}\right\}.
\eea

With this choice, the Lagrangian on the ansatz is therefore modified by $\tilde X^2$. However, another possibility would be 
\be
\tilde X_2^2=\epsilon^{\mu\nu\rho\sigma}\epsilon^{\mu' \nu'\rho'\sigma'}\Tr[L_\mu L_{\mu'}]\Tr[L_\nu L_{\nu'}]\Tr[L_\rho L_{\rho'}]
\Tr[L_\sigma L_{\sigma'}]\;,
\ee
which would vanish on the static ansatz $L_0=0$. With this extension to 3+1 dimensions then, the Lagrangian on the ansatz 
would be unmodified by this term. 

It remains to add the potential (mass) term. 
In chiral perturbation theory, a mass term for one scalar is generalized in the $SU(2)$ case to 
\be
m_\pi^2\phi^2\rightarrow -f_\pi^2m_\pi^2\frac{1}{2}\Tr[U+U^\dagger -2]\;,
\ee
the same that we used in section 4
(the 1/2 is there to cancel the factor of 2 from the trace).\footnote{Note that we can perhaps be more general,  
and also replace a generic potential,
\be
V(\phi)=m_\pi^2f(\phi^2)\rightarrow -m_\pi^2f\left(\frac{1}{2}\Tr[U+U^\dagger-2]\right).
\ee}

Then the $T\bar T$ action in the 3+1 dimensional nonabelian case for $V\neq 0$ will be
(for a potential that is just a mass term, and for $\lambda= 1/(f_\pi^2\b^2)$, 
but note that $F=\phi/f_\pi$ for dimensions to work, where $\phi$ is the usual dimension 1 field)
\bea
{\cal L}^\lambda_{DBI,NLSM}&=&+\frac{f_\pi^2m^2_\pi\Tr(U+U^\dagger-2)}{4\left(1+\frac{m_\pi^2}{2\b^2}\Tr(U+U^\dagger-2)\right)}\cr
&&-f_\pi^2\b^2\frac{\left[\sqrt{1+\frac{1}{4\b^2}\frac{N_c}{2\pi}\left(-\Tr[L_\mu L^\mu]+\frac{N_c}{4\pi}\frac{1}{4\b^2}\tilde X^2\right)
\left(1+\frac{m_\pi^2}{4\b^2}\Tr(U+U^\dagger-2)\right)}-1\right]}{
\left(1+\frac{m_\pi^2}{4\b^2}\Tr(U+U^\dagger-2)\right)}\cr
&&+{\cal L}_{WZW}\;,\label{DBINLSM}
\eea
which is basically the non-Abelian action (\ref{naaction}) rewritten in 3+1 dimensions. 
We refer to this action as a $T\bar T$-like deformation of the Skyrme action. 
It does not fit to the proposal of  \cite{Cardy:2018sdv}  for the deformation in dimensions higher than 1+1.

In analogy to the expansion in terms of small $\lambda$ in eqn (\ref{naaction}) we find now an 
approximate expression for the Lagrangian density
\bea\label{expandthree}
{\cal L}&\simeq& \left(\frac{f_\pi^2}{16}\Tr[L_\mu L^\mu]+
\frac{f_\pi^2m_\pi^2}{4}\Tr[U+U^\dagger -2]\right)\cr
&+&\frac{1}{f_\pi^2\b^2}\left\{\frac{1}{8} \left(\frac{f_\pi^2}{4}\right )^2\left(\frac{N_c}{2\pi}\right)^2\left[ (\Tr[L_\mu L^\mu])^2- 2(\tilde X)^2\right]
 - \left( \frac{f_\pi^2m_\pi^2}{2}\Tr[U+U^\dagger -2] \right)^2    \right.\cr
&&\left.+\frac{1}{4}\left(\frac{f_\pi^2}{4}\right)\left(\frac{N_c}{2\pi}\right)\left(\frac{f_\pi^2m_\pi^2}{2}\Tr[L_\mu L^\mu]
\Tr[U+U^\dagger-2]\right)\right\} 
   + {\cal O}\left(\left(1/(f_\pi^2\b^2)\right )^2\right).
\eea

With respect to the Abelian pion case, we therefore identify the mass potential as
\be
V\leftrightarrow -\frac{m_\pi^2f_\pi^2}{4}\Tr(U+U^\dagger-2)=m_\pi^2 f_\pi^22\sin^2(F/2).\label{pote}
\ee

However, note that $\lambda V=(m_\pi^2/\b^2)2\sin^2(F/2)$, and $m_\pi=134 MeV$ (for $\pi^0$) or $139 MeV$ (for $\pi^\pm$), 
$f_\pi=130 MeV$ (or, depending on convention with respect to multiplication or division by $\sqrt{2}$'s, 
93 MeV or 184 MeV) while $\b$ can only 
be $\sim \Lambda _{\rm QCD}$. Now it depends how this quantity is defined, but if it is defined as 
the string tension, we would expect it to be 
around 0.4 GeV, or perhaps even a factor of 2 smaller. In that case, the factor of $(m_\pi^2/\b^2)2\sin^2 F/2$ can reach 1, 
and the horizon can be achieved.\footnote{However, if one would take as $\b\sim \Lambda_{\rm QCD}$ understood as 
the glueball mass, $\sim 1GeV$, then $1-\lambda V\sim 1-(1/100)2\sin^2F/2$, 
which clearly cannot be =0. That means this choice
doesn't lead to a horizon. In fact, even if we choose $V\sim m_\pi^2 f_\pi^2 F^2/2=m_\pi^2\phi^2/2$, 
that still leads to $1-\lambda V\sim 1-(1/100)F^2$, and 
a topological solution without a horizon has $F(0)=N\pi$ (with $N=1$ for one Skyrmion) and 
$F(\infty)=0$, so $\Delta F=\pi <100$, so no horizon either. 

The only choice, given the parameters, is to have also a $\lambda_0 \phi^4=\lambda_0 f_\pi^4F^4$ 
potential, with $\lambda\lambda_0 f_\pi^4\sim 1/10$ (meaning 
a perturbative potential), so that $(F(0)-F(\infty))^4=\pi^4>10$, giving a horizon. Then we replace in (\ref{DBINLSM}), 
\be
V=-\frac{m_\pi^2f_\pi^2}{4}\Tr(U+U^\dagger-2)\rightarrow -\frac{m_\pi^2 f_\pi^2}{4}
\Tr\left[\log^2\frac{U}{U_0}\right]+\frac{\lambda_0}{2}f_\pi^4\Tr
\left[\log^4\frac{U}{U_0}\right].
\ee

In terms of the Skyrmion ansatz, we have 
\be
V\rightarrow \frac{m_\pi^2}{2}(\phi-\phi_0)^2+\lambda_0(\phi-\phi_0)^4\;,
\ee
so 
\be
1-\lambda V=1-\frac{m_\pi^2}{2\b^2}(F-F_0)^2-\lambda_0\frac{f_\pi^2}{\b^4}(F-F_0)^4\;,
\ee
allowing for the possibility of a horizon. Thus in this case, we would need to consider a rather nontrivial potential, 
but it is also possible to obtain one that has a horizon.}

We can make a comparison with the chiral perturbation theory at order $p^4$ (first nontrivial order). 

If we use $\tilde X_2^2$, this term is of order $p^8$, so doesn't contribute here. If we use the more likely $\tilde X_1^2$, we 
find 
\bea
\tilde X_1^2&=&\left(\Tr[L_\mu L_\mu]\right)^2-\Tr[L_\mu L_\nu]\Tr[L_\mu L_\nu]\Rightarrow\cr
\left(\Tr[L_\mu L^\mu]\right)^2-2\tilde X_1^2&=& -\left(\Tr[L_\mu L_\mu]\right)^2+2\Tr[L_\mu L_\nu]\Tr[L^\mu L^\nu]\cr
&=&-\left(\Tr[\d_\mu U\d^\mu U^\dagger]\right)^2+2\Tr[\d_\mu U \d^\nu U^\dagger]\Tr[\d^\mu U \d^\nu U^\dagger].
\eea

In \cite{Scherer:2002tk}, section 4.7, is written the most 
general Lagrangian at order $p^4$ (like the first nontrivial order in the expansion of the square root for us), in the 
case $m_\pi=0$ (thus with some difference with respect to our case)
\bea
-{\cal L}&=&L_1\left\{\Tr[D_\mu U(D^\mu U)^\dagger]\right\}^2+L_2\Tr\left[D_\mu U(D_\nu U)^\dagger\right]\Tr
\left[D^\mu U(D^\nu U)^\dagger\right]\cr
&&+L_3\Tr\left[D_\mu U(D^\mu U)^\dagger D_\nu U(D^\nu U)^\dagger\right]+L_4\Tr\left[D_\mu U (D^\mu U)^\dagger\right]
\Tr\left(\chi U^\dagger +U \chi^\dagger\right)\cr
&&+L_7\left[\Tr(\chi U^\dagger-U\chi^\dagger)\right]+L_8 \Tr\left(U\chi^\dagger U\chi^\dagger+\chi U^\dagger\chi U^\dagger\right)\cr
&&-iL_9\Tr\left[f_{\mu\nu}^RD^\mu U (D^\nu U)^\dagger+f_{\mu\nu}^L(D^\mu U)^\dagger D^\nu U\right]+L_{10}\Tr\left(Uf_{\mu\nu}^L
U^\dagger f_R^{\mu\nu}\right)\cr
&&+H_1 \Tr\left(f_{\mu\nu}^Rf_R^{\mu\nu}+f_{\mu\nu}^Lf_L^{\mu\nu}\right)+H_2\Tr\left(\chi\chi^\dagger\right).
\eea

The covariant derivatives $D_\mu U$ are related to turning on 
external $a_\mu^L, a_\mu^R$ flavor gauge fields. Needless to say that 
we can also uplift the ordinary derivatives  in (\ref{expandthree}) to covariant ones.  
Besides the terms with scalar sources $\chi$ and external gauge fields $a_\mu^L, a_\mu^R$, there are 10 terms, 
and comparison with experiment is done in their Table 4.3. In it,  the coefficients that are best understood as nonzero
(experimentally, at more than 3$\sigma$ away from zero) are 2,3 and 9,10. 
However, the only terms without external fields are 1, 2 and 3, with $L_1=0.4\pm 0.3$, $L_2=1.35\pm 0.3$ and $L_3=
-3.5\pm 1.1$. We see that expanding the square root in our Lagrangian (\ref{DBINLSM}) we obtain the terms first two terms 
above, with coefficients either 1 and 0 (for $\tilde X_2^2)$, or -1 and +2 (for $\tilde X_1^2$), the last one
consistent with the experimental results above for about 2$\sigma$ in $L_1$ and 2$\sigma$ in $L_2$
(re-normalizing the coefficient of $L_1$ to match), however, the third term would also have
zero coefficient, which is within slightly more than $3\sigma$.Thus, at this point it is  not yet clear if this is consistent with our model. More importantly it is premature to make any claims about the relations between the prediction of our model and the experimental data.




\subsection{Boosted Skyrme-like soliton vs. shockwave solution}

We now consider the boost of the Skyrmion-like solution for the non-Abelian action, and show that 
we get the shockwave solution (in the region where there is 
only one wave). We already showed this in the Abelian analogue case, but now we generalize to the real case (non-Abelian). 

We ignore the potential, since as we saw, the horizon of the Skyrme-like solution will shrink onto 
the singularity $s=0$, so we can safely ignore $V$, and 
consider only the nonabelian DBI of the type considered by Pavlovsky. On the Skyrmion ansatz, this Lagrangian becomes

\be
\sqrt{1+\left(\frac{1}{2\b^2}F'^2+\frac{\sin^2F}{\b^2 r^2}\right)\left(1-\lambda V(F-F_0)\right)}.
\ee
Under the boost, $F'^2$ goes back to $F'^2-\dot F^2$ and, since $r^2\rightarrow \infty$ under $\gamma\rightarrow \infty$
(at finite $x^-$, so finite $s$, meaning for our $\phi(s)$ shockwave), 
the term with $\sin^2F/(\b^2 r^2)$ vanishing. 

That means that on the ansatz, the Lagrangian goes over the the Heisenberg Lagrangian, the 
Abelian scalar DBI one, with $f_\pi(F-F_0)=f_\pi(F-F(r=0))$ identified with the scalar $\phi$
of the abelian action. Then $\phi$ increases away from $r=0$, so $x^-=0$ (with $\tilde r=0$), as does the shockwave solution. 

Then by the logic explained in the case of the Abelian analogue, the Skyrmion-like solution 
should go over to the shockwave one under the infinite boost.

We can also ask whether in the shockwave solution 
we have a baryon-baryon or baryon-antibaryon collision. We saw that the baryon density $\rho_B(r)$ can be calculated, and 
Then, if $F(0)=\pi)$, we have a baryon, and if $F(0)=-\pi$, we have an antibaryon. Moreover, we can see in the density, that for the 
baryon, $F'(r)<0$ ($F$ decreases until $F=0$ at infinity), so $\rho_B(0)>0$, whereas for the antibaryon $F'(r)>0$ ($F$ increases until 
$F=0$ at infinity), so $\rho_B(0)<0$. 

But we also see that, almost by definition, if we have a single, continuous, $F(r)$, or, in the DBI picture, field $\phi(s)$, we have 
a baryon-baryon collision, not a baryon-antibaryon collision, for which $\phi$ (which 
stands in for $F(r)$, up to an additive constant, a we saw)
would have to be discontinuous at the collision point $x=0,t=0$. But we could consider it also to be a baryon-antibaryon solution if we 
could define different $\phi(x=+\infty)$ and $\phi(x=-\infty)$ at fixed $t$. But we can't, since the solution depends on $s=-x^+x^-$ only, 
not independently on $x$ and $t$ (or $x^+$ and $x^-$). This means that the solution does not probe the region $x=\pm \infty$ at 
fixed $t$, in this region the solution would depend on both variables.

\section{Interpretation and discussions}

Finally, we come to the possible interpretation of our results for QCD.

\subsection{Why $T\bar T$?}

The first thing to ask is: why can we use the $T\bar T$ action in the context of  Heisenberg's model? The simplest answer is that 
Heisenberg himself just {\em guessed} an action with the right properties namely, a jump in $(\d_\mu\phi)^2$ for the shockwave 
solution and saturation of the Froissart bound, so we can choose another one that has the same properties and, like we showed, 
has the added property that  it can have both a shockwave and  a soliton solutions, in such a way that the  solitons can be boosted into shockwaves. 

A second point is that the $T\bar T$ deformation action has unique properties, as being the only deformation of 
simple models that can be completely solved, like in the case of integrable models. We could think perhaps of the $T\bar T$ 
deformation of the 3+1 dimensional canonical scalar action (instead of the deformation in 1+1 dimensions). After all, the system 
we are describing is 3+1 dimensional. But we show in Appendix A that this is not possible. We could also think perhaps of the 
$T\bar T$ deformation of the Maxwell or Yang-Mills action for a vector (be it the fundamental gluon, or a low energy vector particle). 
But we show in Appendix B that that is also not possible. 

Accepting this apparent ``uniqueness property", we are still left with the question  how to relate this construction  to the fundamental QCD description 
of the colliding nucleons. 
As we mentioned at the beginning, in the defining relation of the $T\bar T$ deformation Lagrangian (\ref{equ})
all objects are quantum renormalized and UV finite, and on the right-hand side we have an operator regularized by point-splitting. 
Yet we treated the resulting Lagrangian as a classical one, giving rise to classical fields, which is in accordance with the reasoning 
of Heisenberg for his model. 

The simplest possibility of relating the $T\bar T$ deformed Lagrangian to QCD then is to think of it as being a quantum effective 
action in the Wilsonian sense for the pions (or rather, a single (pseudo)scalar version of it, that we tried to extend to $SU(2)$-valued 
fields). The parameter $\lambda$ of the deformation (corresponding to $l^4$ for the usual Heisenberg model (\ref{Heisaction}))
{\em must be related to the QCD string tension}.

The quantum theory for this Wilsonian  effective theory, now reduced 
(the Wilsonian theory, not its quantum theory)
to 1+1 dimensions (the direction of propagation and time) should give the defining equation of the $T\bar T$ deformation
(\ref{equ}), as a function of $\lambda$.  
Note that this 
would not mean that the same equation (\ref{equ})
is satisfied for the full quantum theory of the 3+1 dimensional Wilsonian effective theory (in fact, we saw in Appendix A that the 
$T\bar T$ deformed 3+1 dimensional theory doesn't have the needed classical solutions).

The defining relation (\ref{equ}) can be rewritten, with $\lambda$= the string tension $\sigma$, 
\be
\frac{\d {\cal L}^\sigma}{\d \sigma}=\det T_{\mu\nu}^{(\sigma)}.
\ee

One possible interpretation of this relation
is that as we vary the string tension, the variation of this Wilson effective Lagrangian is proportional to the density square of the 
pion field, since the pion field breaks up into a quark at one end and an anti-quark at the other, but both are proportional to the density 
at that point (note that this is how we understand the defining relation anyway: by point splitting). And then 
the Lagrangian changes by the variation of this pion particle energy density.

The Wilsonian effective theory reduced to 1+1 dimensions admits a classical limit as well, 
which is when the value of its field is large (many 
pions in the same state), so we can talk about a classical profile. 
For 1+1 dimensions, we have concentrated on the shockwave, which is a statement 
about the field around a nucleus, moving at high speed. 
There is a legitimate question as to why the Wilsonian theory is still valid, if we are at 
high energies, but perhaps the answer is simply that we consider the field around a nucleus, 
and the nucleus has high energy, yet the pions do not, 
since they are {\em virtual} (part of the field). Indeed, the DBI action is used for the saturation of the 
Froissart bound for $\sigma_{\rm tot}$, which 
by the optical theorem is $\sim $ Im ${\cal M}(\tilde s,t=0)<\sim  \tilde s\ln^2 \tilde s$, (see also around 
eq. 28 in section 4 of \cite{Adams:2006sv})
which are both dominated by emission of large number of soft (low energy) pions, which would make a 
classical (many particles), but low energy field.

\subsection{Gravity dual interpretation}

Like in \cite{Nastase:2015ixa}, we can ask what is the interpretation from a gravity dual point of 
view? As usual, the $\phi$ scalar is the position of some D-brane in the fifth dimension. But we have 
already analyzed this in \cite{Nastase:2015ixa}, even for the nonlinear sigma model
$g_{ij}(\phi)\d_\mu \phi^i\d^\mu \phi^j$, in eq. (4.15) there. Because of eq. (4.18) there, 
we have a single $g(\phi)$, and by comparison with our case (\ref{Lagrang}), we have
\be
g(\phi)=\bar\lambda(\phi)=\lambda(1-\lambda V(\phi)).
\ee

We had analyzed in detail the case $g(\phi)=1/\phi^4$, which corresponds to motion in AdS space 
in the gravity dual. But here, we see that the metric in the fifth dimension is dependent of $V(\phi)$, 
and moreover has a kind of ``horizon'', but only in the sense that the metric $g(\phi)$ 
changes sign at $\phi_0$.

Until now, we have compared with the our holographic model, eq. (4.15) in  \cite{Nastase:2015ixa}, 
only in the metric $g(\phi)$. But the $T\bar T$ action has more terms. In particular, 
there is an $e^{-\varphi}$ factor (where $\varphi$ is the dilaton)
multiplying the square root, just like $1/\bar \lambda$ in (\ref{Lagrang}), 
so we identify those two. Then there is another term, outside the square root, $-\tilde V$, that 
is just like the WZ term $A_{0123}e^{\varphi}$, so again we must identify the two, leading to 
\bea
g(\phi)&=&\bar\lambda=\lambda(1-\lambda V(\phi))\cr
T_pe^{\varphi(\phi)}&=& \bar \lambda =\lambda(1-\lambda V(\phi))\cr
\mu_pA_{0123}e^{\varphi}(\phi)&=&-\tilde V(\phi)=\frac{1-2\lambda V(\phi)}{2\lambda (1-\lambda V(\phi))}.
\eea

Note that since the dilaton $\varphi(\phi)$ must be positive, and since at the "horizon" $1-\lambda V=0$, $T_p e^{-\varphi(\phi)}$ changes 
sign, it follows that the tension $T_p$ changes sign, from D-brane type, to orientifold-type. However, a orientifold does not fluctuate like a 
D-brane, so it is a negative tension brane in the sense of Randall-Sundrum (which however doesn't have a string theory interpretation). 
This is then a truly phenomenological interpretation of AdS/CFT: just use the map, without worrying about string theory rules and derivations. 
Yet, note that the since at this "horizon", the brane becomes tensionless, so states are massless, our probe interpretation for the D-brane 
becomes meaningless, hence the singularity is a fake one, that needs to be resolved. What we do have instead is a phase transition
in the string-like description (between D-branes and "orientifold-like objects"). In terms of the field theory side, 
the Wilsonian description in terms of only pions is no longer complete, 
we have to introduce the nucleons as well. We "cheat it in" by using the Skyrmion-like solution for the pion field to describe the nucleons.

\section{Conclusions and open questions}


In this paper we have analyzed  actions of   $T\bar T$ deformations of the actions of  canonical scalars. 
The purpose  was to   use them as     effective actions, both 
in the Heisenberg model for high energy nucleon scattering, and in a Skyrme-like action that admits (topological) solitons that can be 
identified with the nucleons themselves. 

We have first found that for the 1+1 dimensional $T\bar T$ deformation of a canonical scalar, which we have found in a 
companion paper \cite{HNJS} (see also \cite{Conti:2018jho,Conti:2018tca}) that it  has soliton solutions, as well as shockwave 
solutions, the latter can be used  to describe nucleons and their scatterings that    saturate of the Froissart bound. Moreover, we can 
understand the shockwave solution as corresponding to the collision of two boosted solitons. 

We have found then that the same action, when extended to 3+1 dimensions, has both shockwave solutions and
BIon-like solutions, with finite radial derivative, 
while the Heisenberg action (\ref{Heisaction}) has catenoid solutions, with horizons where the radial derivative blows up. We have 
argued that the boosted BIon-like solutions should describe the shockwave solutions. 

The 3+1 dimensional nonabelian scalar version of the $T\bar T$ deformation action, was found to have Skyrmion-like solutions
and shockwave solutions describing the collision of two boosted Skyrmion-like solutions. It was also found to be possibly consistent with 
the phenomenological fit of chiral perturbation theory to experiment.

We interpreted these results as saying that the $T\bar T$ deformation action should be understood as 
a quantum Wilsonian effective action for QCD, reduced to 1+1 dimensions, though why do we have this equality is unclear. 
We also attempted an interpretation from the point of view of a gravity dual, but the interpretation could not be completed.

There are plenty of open questions that are awaiting further investigation. Here we list some of them:
\begin{itemize}
\item
The $T\bar T$ deformation  at higher than 1+1 dimensions is not unique. 
In \cite{Cardy:2018sdv} it was proposed to use  $|det {T}|^{\frac1\alpha}$ with $\alpha=d-1$ and in \cite{Bonelli:2018kik} 
it was further generalized to different values of $\alpha$. An interesting question is what is the deformation condition 
that yields the $T\bar T$-like deformation that we are using (\ref{DBINL}). 
\item
The relation between $T\bar T$ deformed actions and QCD.  We view the current paper as just a starting point for this analysis.

\item
The standard theory of Skyrmions has been plagued with several 
significant  problems. In particular (i) The results show that if the pion
mass is set to its experimental value then the nucleon and delta masses can not be
reproduced for any values of the Skyrme parameters; the commonly used Skyrme
parameters are simply an artifact of the rigid body approximation\cite{Battye:2005nx}. (ii)  It yields nuclear binding 
energies that are an order of magnitude larger
than experimental nuclear data $15\%$ instead of $1\%$\cite{Naya:2018kyi}.  (iii) It predicts intrinsic shapes for nuclei that fail to match the
clustering structure of light nuclei.
It will be very interesting to explore the whether   these problems  can be circumvented  in the deformed Skyrme theory. 
\item
Expanding the deformed Skyrme acion in terms of the deformation parameter yields the Skyrme action plus 
additional terms that are higher order powers in the derivative of the fields. Such terms  are part of the chiral 
Lagrangian. Thus, it will be very interesting  to perform a  comprehensive comparison between the coefficients 
of these terms that determined by  the experimental data and those that follow from the expansion of the deformed action.  
\end{itemize}

\section*{Acknowledgements}


We thank Aki Hashimoto 
for useful discussions. The work of HN is supported in part by CNPq grant 301491/2019-4 and FAPESP grants 2019/21281-4 
and 2019/13231-7. HN would also like to thank the ICTP-SAIFR for their support through FAPESP grant 2016/01343-7.
The work of J.S   was supported in part by a center of excellence funded  by the Israel Science Foundation (grant number 2289/18).

\appendix

\section{3+1 dimensional $T\bar T$ deformation and the Heisenberg model}

In this Appendix we will consider one possible $T\bar T$ deformation of the free scalar in 3+1 dimensions, defined in \cite{Bonelli:2018kik}, 
and we will analyze its applicability to the Heisenberg model. 

The deformed Lagrangian in the presence of a general potential $V$ was defined implicitly, as a solution of an equation, which is impossible 
to solve in a (simple) closed form in the general case. This would be needed for $V=m^2\phi^2/2$, 
for instance to analyze the radiation arising from a possible shockwave solution, as seen in the text. 

However, we will content ourselves with the case of $V=0$ ($m=0$), since this suffices in order to find the leading order shockwave solution 
near $s=0$, where any mass scale is irrelevant. In that case, the equation for the Lagrangian can be solved exactly for any spacetime 
dimension $D$, with result
\be
-{\cal L}_{D,1}(t,X)=\left\{ \frac{1}{2t}\left[\sqrt{1+4t\left[\frac{(\d_\mu \phi)^2}{2}\right]^{D-1}}-1\right]\right\}^{\frac{1}{D-1}}.
\ee

Consider, as in the text, a solution $\phi(s)$ depending only on $s=t^2-x^2$, and independent also of $y,z$. If the solution is a shockwave, 
then moreover $\phi(s)=0$ for $s<0$. Then 
\be
X\equiv (\d_\mu \phi)^2=-4s \left(\frac{d\phi}{ds}\right)^2\;,
\ee
so the Lagrangian on the shockwave solution is 
\be
-{\cal L}_{D,1}(t,X)=\left\{\frac{1}{2t}\left[\sqrt{1+4t\left[-\frac{4s}{2}\left(\frac{d\phi}{ds}\right)^2\right]^{D-1}}-1\right]\right\}^{\frac{1}{D-1}}.
\ee

The equations of motion are 
\bea
&&\frac{d}{ds}\left\{
\left\{\frac{1}{2t}\left[\sqrt{1+4t\left[-\frac{4s}{2}\left(\frac{d\phi}{ds}\right)^2\right]^{D-1}}-1\right]\right\}^{\frac{2-D}{D-1}}
\left\{\frac{1}{2t}\frac{4t(D-1)}{\sqrt{1+4t\left[-\frac{4s}{2}\left(\frac{d\phi}{ds}\right)^2\right]^{D-1}}}\times\right.\right.\cr
&&\left.\left.\times \left[-2s\left(\frac{d\phi}{ds}\right)^2\right]^{D-2}\left(-4s\frac{d\phi}{ds}\right)\right\}\right\}=0.\label{ttdeom}
\eea

Define
\be
\frac{4t}{2^{D-1}}\equiv l^{4(D-1)}\;,
\ee
so that the Lagrangian on the shockwave solution reads
\be
-{\cal L}_{D,1}(t,X)=\left\{\frac{1}{2t}\left[\sqrt{1+\left[-l^44s\left(\frac{d\phi}{ds}\right)^2\right]^{D-1}}-1\right]\right\}^{\frac{1}{D-1}}\;,
\ee
and the equations of motion read
\bea
&&\frac{d}{ds}\left\{
\left\{\left[\sqrt{1+\left[-l^44s\left(\frac{d\phi}{ds}\right)^2\right]^{D-1}}-1\right]\right\}^{\frac{2-D}{D-1}}
\left\{\frac{(D-1)}{\sqrt{1+\left[-l^44s\left(\frac{d\phi}{ds}\right)^2\right]^{D-1}}}\times\right.\right.\cr
&&\left.\left.\times \left[-4s\left(\frac{d\phi}{ds}\right)^2\right]^{D-2}\left(-8s\frac{d\phi}{ds}\right)\right\}\right\}=0.\label{ttdeomp}
\eea

The we check whether Heisenberg's near $s=0$ solution 
\be
\phi(s)=l^{-2}\sqrt{s}\label{soluHeis}
\ee
is still valid for the equations of motion (\ref{ttdeomp}) for the $D$ dimensional Lagrangian. We first note that, if it is, then the square root 
in the Lagrangian and equations of motion still vanishes. 

One can check explicitly, that in the Heisenberg case, corresponding to $D=2$, the solution comes from the leading terms, coming from 
derivatives of the square root, and in the end, the leading terms in the equation of motion on the solution reduce to 
\be
\frac{d}{ds}\left[s\left(\frac{d\phi}{ds}\right)^2\right]=0\;,
\ee
which is still true in the $D$ dimensional case. Moreover, like in the $D=2$ case, the leading terms, coming from the derivatives of the 
square root, will also result in the above equation. It is then clear that (\ref{soluHeis}) continues to be a solution in the $D$ dimensional 
case. 

However, in that case, the on-shell Lagrangian for the shockwave solution is 
\be
{\cal L}_{\rm on-shell}=(-1)^{\frac{1}{D-1}}=e^{\frac{2\pi i}{D}}\in \mathbb{C}.
\ee
This is only real in the $D=2$ case, but the Lagrangian is supposed to be real on classical solutions. The only possible exceptions 
are quantum solutions like instantons, describing some transition from the quantum path integral. 

Since the Lagrangian is supposed to be an effective one, describing a regime of low energy, it cannot be used to describe quantum transitions, 
so there is no possibility to use for for $D\neq 2$, in particular cannot be used for $D=4$. 

\section{On the Abelian and non-Abelian $T\bar T$ deformations of Maxwell and Yang-Mills theories}

In this Appendix we consider the Abelian and non-Abelian $T\bar T$ deformations of Maxwell and Yang-Mills theories
in \cite{Brennan:2019azg}, and find that 
the BIon solution of Born and Infeld is not a solution, and moreover, there is no analogue of this soliton solution for these theories. 

The Lagrangian is given in an implicit form, in terms of 
\be
\chi\equiv\lambda \Tr F_{\mu\nu} F^{\mu\nu}\;,\;\;\; \lambda {\cal L}\equiv h(\chi)
\ee
by the equation
\be
\chi =\frac{1}{256} (1-\sqrt{1-8h(\chi)})(3+\sqrt{1-8h(\chi)}).\label{ttbareq}
\ee

By applying $d/d\chi$ on the above equation, we obtain 
\be
1\leq \frac{dh}{d\chi}=\frac{16}{(3+\sqrt{1-8h})^2}\leq \frac{16}{9}.
\ee

On the other hand, it is easy to see that the equation of motion (with respect to the gauge field $A_\mu$) is 
\be
D_\mu \left(\frac{dh}{d\chi}F_{\mu\nu}\right)=D_\mu \left(\frac{16}{(3+\sqrt{1-8h})^2}F_{\mu\nu}\right)=0.
\ee

Another way to think about this is to defined, as in the main text (and as in a medium), 
\be
\frac{\d h}{\d E_i}\equiv D_i\;,
\ee
such that the equation of motion for a static solution is 
\be
\d_i D_i=\vec{\nabla}\cdot \vec{D}=0\;,
\ee
or rather, $q\delta^3(r)$ on the right-hand side, for a delta function source. Here 
\be
\vec{D}=\frac{dh}{d\chi}\vec{E}.\label{dhe}
\ee

On the other hand, $\vec{E}$ remains finite, just like in the usual Born-Infeld theory. Indeed, in 2 dimensions, 
\cite{Brennan:2019azg} show that, defining $F^2=F_{\mu\nu} F^{\mu\nu}=-2 F_{01}^2=-2 E^2$ in Minkowski space, the Born-Infeld 
Lagrangian is
\be
{\cal L}=\sqrt{1+4\lambda F^2}=\sqrt{1-8\lambda F_{01}^2}\;,
\ee
so that 
\be
E=F_{01}\leq \frac{1}{\sqrt{8\lambda}}.
\ee

For the $T\bar T$ theory, maximizing the right-hand side of (\ref{ttbareq}), which happens at ${\cal L}=1/(8\lambda)$, where the
right-hand side takes the value 27/256, so that 
\be
F^2\leq \frac{27}{256\lambda}\Rightarrow E=F_{01}\leq \sqrt{\frac{27}{512\lambda}}\;,
\ee
which is just very slightly smaller than the maximum value for Born-Infeld. 

Coming back to our issue, we see that $D_i$ remains finite, since both terms on the right-hand side of (\ref{dhe}) remain finite. 
This is unlike the case of the Born-Infeld soliton, where $\vec{E}\leq 1$ (in units of $E_{\rm max}=1/\sqrt{8\lambda}$ here), but 
\be
\vec{D}=-\frac{\vec{E}}{\sqrt{1-\vec{E}^2}}\rightarrow 0\;,
\ee
the singularity coming from it being the singular solution of the $\vec{\nabla}\cdot\vec{D}=\delta^2(r)$ equation. 
This was found first in 3+1 dimensions, but the same 
continues to be valid in 1+1 dimensions, where for Born-Infeld we have
\be
D=\frac{\d {\cal L}}{\d E}=-\frac{E}{\sqrt{1-8\lambda E^2}}\;,
\ee
and again the singularity comes from it being a solution of the $\d D=\delta(r)$ equation. 

In conclusion, since $\vec{D}$ is non-singular for the case of the $T\bar T$ action, we cannot have BI-type soliton solutions for it.


\bibliography{DBITTbar}
\bibliographystyle{utphys}

\end{document}